\documentclass[superscriptaddress]{revtex4}
\usepackage{amsfonts}
\usepackage{amsmath}
\usepackage{amssymb}
\usepackage{graphics}

\input{tcilatex}
\begin{document}

\author{Mario Castagnino}
\affiliation{CONICET-IAFE-IFIR-Universidad de Buenos Aires, Argentina}
\author{Sebastian Fortin}
\affiliation{CONICET-IAFE-Universidad de Buenos Aires, Argentina}
\author{Olimpia Lombardi}
\affiliation{CONICET-Universidad de Buenos Aires, Argentina}

\begin{abstract}
The works on decoherence due to spin baths usually agree in studying a
one-spin system in interaction with a large spin bath. In this paper we
generalize those models by analyzing a many-spin system and by studying
decoherence or its suppression in function of the relation between the
numbers of spins of the system and the bath. This model may help to identify
clusters of particles unaffected by decoherence, which, as a consequence,
can be used to store quantum information.
\end{abstract}

\title{Suppression of decoherence in a generalization of the spin-bath model}
\pacs{03.65.Yz 03.67.Bg 03.67.Mn 03.65.Ud}
\keywords{Decoherence, closed-system, spin-bath}
\maketitle

\section{Introduction}

Decoherence\ refers to the quantum process that turns a coherent pure state
into a decohered mixed state. It is essential in the account of the
emergence of classicality from quantum behavior, since it explains how
interference vanishes in an extremely short decoherence time. The orthodox
explanation of the phenomenon is given by the \textit{environment-induced
decoherence} approach (see Refs. \cite{Zurek-1982}, \cite{Zurek-1993}, \cite%
{Paz-Zurek}, \cite{Zurek-2003}), according to which decoherence is a process
resulting from the interaction of an open quantum system and its
environment. By studying different physical models, it is proved that the
reduced state $\rho _{S}(t)=Tr_{E}\rho _{SE}(t)$ of the open system rapidly
diagonalizes in a well defined pointer basis, which identifies the
candidates for classical states.

The environment-induced approach has been extensively applied to many areas
of physics $-$such as atomic physics, quantum optics and condensed matter$-$%
, and has acquired a great importance in quantum computation, where the loss
of coherence represents a major difficulty for the implementation of
information processing hardware that takes advantage of superpositions. In
particular, decoherence resulting from the interaction with nuclear spins is
the main obstacle to quantum computations in magnetic systems. This fact has
lead to a growing interest in the study of decoherence due to spin baths
(see Refs. \cite{Paganelli} to \cite{Lombardo}). By beginning from the
seminal paper of Zurek (\cite{Zurek-1982}), many works have studied the
decoherence due to a collection of independent spins. More recently, some
papers have directed the attention to the interactions between modes within
the bath. For instance, by studying a central spin coupled to a spin-bath,
Tessieri and Wilkie (\cite{Tessieri}) showed that, whereas in the absence of
intra-environmental coupling the decoherence of the central spin is fast and
irreversible, strong intra-environmental coupling leads to decoherence
suppression. The same model was further analyzed by Dawson \textit{et al.} (%
\cite{Dawson}), with the purpose of relating decoherence with the pairwise
entanglement between individual bath spins. In turn, Rossini \textit{et al.}
(\cite{Rossini}) left behind the assumption that the central spin is coupled
isotropically to all the spins of the bath, and considered the case where
the spin system interacts with only few spins of the bath.

Our analysis can be framed in the context of the above works; it aims at
generalizing the paradigmatic spin-bath model. In fact, most of the works
done so far agree in studying a one-spin system in interaction with a large
spin bath. The crucial feature of our work is the analysis of a many-spin
system, and the study of decoherence or its suppression in function of the
relation between the numbers of spins of the system and the bath. This
generalized spin-bath model can also be conceived as a partition of a whole
closed system into an open many-spin system and its environment. From this
perspective, we can study different partitions of the whole system and
identify those for which the selected system does not decohere; this might
allow to define clusters of particles that can be used to store q-bits.

In order to develop our analysis, we will rely on the general framework for
decoherence introduced in Ref. \cite{CFL}, where the split of a closed
quantum system into an open subsystem and its environment is just conceived
as a way of selecting a particular space of relevant observables of the
whole closed system. Since there are many different spaces of relevant
observables depending on the observational viewpoint adopted, the same
closed system can be decomposed in many different ways: each decomposition
represents a decision about which degrees of freedom are relevant and which
can be disregarded in each case.

On this basis, the paper is organized as follows. In Section II, the
standard spin-bath model is presented from the general framework
perspective: this presentation will allow us to consider two different
decompositions, which supply the basis for comparing the results obtained
for the generalized model in the following sections.\ In Sections III, IV
and V, the generalization of the spin-bath model is presented and solved by
computer simulations; this task will allow us to compare the results
obtained for two different ways of splitting the entire closed system into
an open system and its environment.\ Finally, in Section VI we introduce our
concluding remarks.

\section{The spin-bath model}

The spin-bath model is a very simple model that has been exactly solved in
previous papers (see \cite{Zurek-1982}). Here we will recall its main
results, obtained from the general framework introduced in Ref. \cite{CFL},
in order to compare the analogous results to be obtained in the next
sections for the generalized model.

\subsection{Presentation of the model}

Let us consider a closed system $U=P\cup P_{1}\cup \ldots \cup P_{N}=P\cup
(\cup _{i=1}^{N}P_{i})$, where (i) $P$ is a spin-1/2 particle represented in
the Hilbert space $\mathcal{H}_{P}$, and (ii) each $P_{i}$ is a spin-1/2
particle represented in its Hilbert space $\mathcal{H}_{i}$. The Hilbert
space of the composite system $U$\ is, then, 
\begin{equation}
\mathcal{H}=\mathcal{H}_{P}\otimes \left( \bigotimes\limits_{i=1}^{N}%
\mathcal{H}_{i}\right)  \label{3.0}
\end{equation}%
In the particle $P$, the two eigenstates of the spin operator $S_{P,%
\overrightarrow{v}}$\ in direction $\overrightarrow{v}$ are $\left\vert
\Uparrow \right\rangle ,\left\vert \Downarrow \right\rangle $, such that $%
S_{P,\overrightarrow{v}}\left\vert \Uparrow \right\rangle =\frac{1}{2}%
\left\vert \Uparrow \right\rangle $ and $S_{P,\overrightarrow{v}}\left\vert
\Downarrow \right\rangle =-\frac{1}{2}\left\vert \Downarrow \right\rangle $.
In each particle $P_{i}$, the two eigenstates of the corresponding spin
operator $S_{i,\overrightarrow{v}}$\ in direction $\overrightarrow{v}$ are $%
\left\vert \uparrow _{i}\right\rangle ,\left\vert \downarrow
_{i}\right\rangle $, such that $S_{i,\overrightarrow{v}}\left\vert \uparrow
_{i}\right\rangle =\frac{1}{2}\left\vert \uparrow _{i}\right\rangle $ and $%
S_{i,\overrightarrow{v}}\left\vert \downarrow _{i}\right\rangle =-\frac{1}{2}%
\left\vert \downarrow _{i}\right\rangle $. Therefore, a pure initial state
of $U$ reads%
\begin{equation}
|\psi _{0}\rangle =(a\left\vert \Uparrow \right\rangle +b\left\vert
\Downarrow \right\rangle )\otimes \left( \bigotimes_{i=1}^{N}(\alpha
_{i}|\uparrow _{i}\rangle +\beta _{i}|\downarrow _{i}\rangle )\right)
\label{3.3}
\end{equation}%
where $\left\vert a\right\vert ^{2}+\left\vert b\right\vert ^{2}=1$ and $%
\left\vert \alpha _{i}\right\vert ^{2}+\left\vert \beta _{i}\right\vert
^{2}=1$. If the self-Hamiltonians $H_{P}$ of $P$ and $H_{i}$ of $P_{i}$ are
taken to be zero, and there is no interaction among the $P_{i}$, then the
total Hamiltonian $H$ of the composite system $U$ is given by the
interaction between the particle $P$ and each particle $P_{i}$ (see \cite%
{Zurek-1982}, \cite{Max}):%
\begin{equation}
H=\frac{1}{2}\left( \left\vert \Uparrow \right\rangle \left\langle \Uparrow
\right\vert -\left\vert \Downarrow \right\rangle \left\langle \Downarrow
\right\vert \right) \otimes \sum_{i=1}^{N}\left[ g_{i}\left( \left\vert
\uparrow _{i}\right\rangle \left\langle \uparrow _{i}\right\vert -\left\vert
\downarrow _{i}\right\rangle \left\langle \downarrow _{i}\right\vert \right)
\otimes \left( \bigotimes_{j\neq i}^{N}\mathbb{I}_{j}\right) \right]
\label{3.4}
\end{equation}%
where $\mathbb{I}_{j}=\left\vert \uparrow _{j}\right\rangle \left\langle
\uparrow _{j}\right\vert +\left\vert \downarrow _{j}\right\rangle
\left\langle \downarrow _{j}\right\vert $ is the identity operator on the
subspace $\mathcal{H}_{j}$. Under the action of $H$, the state $|\psi
_{0}\rangle $ evolves into $\left\vert \psi (t)\right\rangle =a\left\vert
\Uparrow \right\rangle |\mathcal{E}_{\Uparrow }(t)\rangle +b\left\vert
\Downarrow \right\rangle |\mathcal{E}_{\Downarrow }(t)\rangle $ where 
\begin{equation}
\left\vert \mathcal{E}_{\Uparrow }(t)\right\rangle =\left\vert \mathcal{E}%
_{\Downarrow }(-t)\right\rangle =\bigotimes_{i=1}^{N}\left( \alpha
_{i}\,e^{-ig_{i}t/2}\,\left\vert \uparrow _{i}\right\rangle +\beta
_{i}\,e^{ig_{i}t/2}\,\left\vert \downarrow _{i}\right\rangle \right)
\label{3.6}
\end{equation}

The space $\mathcal{O}$ of the observables of the composite system $U$ can
be obtained as $\mathcal{O}=\mathcal{O}_{P}\otimes (\otimes _{i=1}^{N}%
\mathcal{O}_{i})$, where $\mathcal{O}_{P}$ is the space of the observables
of the particle $P$ and $\mathcal{O}_{i}$ is the space of the observables of
the particle $P_{i}$. Then, an observable $O\in \mathcal{O}=\mathcal{H}%
\otimes \mathcal{H}$ can be expressed as%
\begin{equation}
O=O_{P}\otimes (\bigotimes_{i=1}^{N}O_{i})  \label{3.7.1}
\end{equation}%
where%
\begin{eqnarray}
O_{P} &=&s_{\Uparrow \Uparrow }\left\vert \Uparrow \right\rangle
\left\langle \Uparrow \right\vert +s_{\Uparrow \Downarrow }\left\vert
\Uparrow \right\rangle \left\langle \Downarrow \right\vert +s_{\Downarrow
\Uparrow }\left\vert \Downarrow \right\rangle \left\langle \Uparrow
\right\vert +s_{\Downarrow \Downarrow }\left\vert \Downarrow \right\rangle
\left\langle \Downarrow \right\vert \ \in \mathcal{O}_{P}  \label{3.7.2} \\
O_{i} &=&\epsilon _{\uparrow \uparrow }^{(i)}|\uparrow _{i}\rangle \langle
\uparrow _{i}|+\epsilon _{\downarrow \downarrow }^{(i)}|\downarrow
_{i}\rangle \langle \downarrow _{i}|+\epsilon _{\downarrow \uparrow
}^{(i)}|\downarrow _{i}\rangle \langle \uparrow _{i}|+\epsilon _{\uparrow
\downarrow }^{(i)}|\uparrow _{i}\rangle \langle \downarrow _{i}|\ \in 
\mathcal{O}_{i}  \label{3.7.3}
\end{eqnarray}%
Since the operators $O_{P}$ and $O_{i}$ are Hermitian, the diagonal
components $s_{\Uparrow \Uparrow }$, $s_{\Downarrow \Downarrow }$, $\epsilon
_{\uparrow \uparrow }^{(i)}$, $\epsilon _{\downarrow \downarrow }^{(i)}$ are
real numbers, and the off-diagonal components are complex numbers satisfying 
$s_{\Uparrow \Downarrow }=s_{\Downarrow \Uparrow }^{\ast }$, $\epsilon
_{\uparrow \downarrow }^{(i)}=\epsilon _{\downarrow \uparrow }^{(i)\ast }$.
Then, the expectation value of the observable $O$ in the state $\left\vert
\psi (t)\right\rangle $ can be computed as%
\begin{equation}
\langle O\rangle _{\psi (t)}=(|a|^{2}s_{\Uparrow \Uparrow
}+|b|^{2}s_{\Downarrow \Downarrow })\,\Gamma _{0}(t)+2\func{Re}\,[ab^{\ast
}\,s_{\Downarrow \Uparrow }\,\Gamma _{1}(t)]  \label{3.8}
\end{equation}%
where (see \cite{Max}) 
\begin{eqnarray}
\Gamma _{0}(t) &=&\prod_{i=1}^{N}\left[ |\alpha _{i}|^{2}\epsilon _{\uparrow
\uparrow }^{(i)}+|\beta _{i}|^{2}\epsilon _{\downarrow \downarrow }^{(i)}+2%
\func{Re}(\alpha _{i}{}\,\beta _{i}^{\ast }\epsilon _{\downarrow \uparrow
}^{(i)}e^{ig_{i}t})\right]  \label{3.9} \\
\Gamma _{1}(t) &=&\prod_{i=1}^{N}\left[ |\alpha _{i}|^{2}\epsilon _{\uparrow
\uparrow }^{(i)}e^{ig_{i}t}+|\beta _{i}|^{2}\epsilon _{\downarrow \downarrow
}^{(i)}e^{-ig_{i}t}+2\func{Re}(\alpha _{i}{}\,\beta _{i}^{\ast }\epsilon
_{\downarrow \uparrow }^{(i)})\right]  \label{3.10}
\end{eqnarray}%
By contrast to the usual presentations, we will study two different
decompositions of the whole closed system $U$ into a relevant part and its
environment.

\subsection{The spin-bath model: Decomposition 1}

In the typical presentations of the model, the open system $S$ is the
particle $P$, and the remaining particles $P_{i}$ play the role of the
environment $E$: $S=P$ and $E=\cup _{i=1}^{N}P_{i}$. Then, the Hilbert space
decomposition for this case is 
\begin{equation}
\mathcal{H}=\mathcal{H}_{S}\otimes \mathcal{H}_{E}=\left( \mathcal{H}%
_{P}\right) \otimes \left( \bigotimes\limits_{i=1}^{N}\mathcal{H}_{i}\right)
\label{3.10-1}
\end{equation}%
Therefore, the relevant observables $O_{R}$ of the closed system $U$ are
those corresponding to the particle $P$, and they are obtained from eqs. (%
\ref{3.7.1}), (\ref{3.7.2}) and (\ref{3.7.3}), by making $\epsilon
_{\uparrow \uparrow }^{(i)}=\epsilon _{\downarrow \downarrow }^{(i)}=1$ and $%
\epsilon _{\uparrow \downarrow }^{(i)}=0$:%
\begin{equation}
O_{R}=O_{S}\otimes \mathbb{I}_{E}=\left( \sum_{s,s^{\prime }=\Uparrow
,\Downarrow }s_{ss^{\prime }}|s\rangle \langle s^{\prime }|\right) \otimes
\left( \bigotimes_{i=1}^{N}\mathbb{I}_{i}\right)  \label{3.11}
\end{equation}%
The expectation value of these observables in the state $\left\vert \psi
(t)\right\rangle $ is given by%
\begin{equation}
\langle O_{R}\rangle _{\psi (t)}=|a|^{2}\,s_{\Uparrow \Uparrow
}+|b|^{2}\,s_{\Downarrow \Downarrow }+2\func{Re}[ab^{\ast }\,s_{\Downarrow
\Uparrow }\,r(t)]  \label{3.12}
\end{equation}%
where 
\begin{equation}
r(t)=\langle \mathcal{E}_{\Downarrow }(t)\rangle |\mathcal{E}_{\Uparrow
}(t)\rangle =\prod_{i=1}^{N}\left( |\alpha _{i}|^{2}\,e^{-ig_{i}t}+|\beta
_{i}|^{2}\,e^{ig_{i}t}\right)  \label{3.13}
\end{equation}%
and, then, 
\begin{equation}
|r(t)|^{2}=\prod_{i=1}^{N}(|\alpha _{i}|^{4}+|\beta _{i}|^{4}+2|\alpha
_{i}|^{2}|\beta _{i}|^{2}\cos 2g_{i}t)  \label{3.14}
\end{equation}%
This means that, in eq. (\ref{3.8}), $\Gamma _{0}(t)=1$ and $\Gamma
_{1}(t)=r(t)$.

If we take $\left\vert \alpha _{i}\right\vert ^{2}$ and $\left\vert \beta
_{i}\right\vert ^{2}$ as random numbers in the closed interval $\left[ 0,1%
\right] $, then $|r(t)|^{2}$ is an infinite product of numbers belonging to
the open interval $\left( 0,1\right) $. As a consequence, $%
\lim_{N\rightarrow \infty }r(t)=0$. Therefore, it can be expected that, for $%
N$ finite, $r(t)$ will evolve in time from $r(0)=1$ to a very small value
(see numerical simulations in \cite{Zurek-1982} and \cite{Max}).

\subsection{The spin-bath model: Decomposition 2}

Although in the usual presentations of the model the open system of interest
is $P$, we can conceive different ways of splitting the whole closed system $%
U$ into an open system $S$ and its environment $E$. For instance, we can
decide to observe a particular particle $P_{j}$ of what was previously
considered the environment, and to consider the remaining particles as the
new environment, in such a way that $S=P_{j}$ and $E=P\cup (\cup _{i=1,i\neq
j}^{N}P_{i})$. The total Hilbert space of the closed composite system $U$\
is still given by eq. (\ref{3.0}), but in this case the corresponding
decomposition is 
\begin{equation}
\mathcal{H}=\mathcal{H}_{S}\otimes \mathcal{H}_{E}=\left( \mathcal{H}%
_{j}\right) \otimes \left( \mathcal{H}_{P}\otimes \left( \bigotimes\limits 
_{\substack{ i=1  \\ i\neq j}}^{N}\mathcal{H}_{i}\right) \right)  \label{4-0}
\end{equation}%
and the relevant observables $O_{R}$ of the closed system $U$ are those
corresponding to the particle $P_{j}$:%
\begin{equation}
O_{R}=O_{S}\otimes \mathbb{I}_{E}=O_{P_{j}}\otimes \left( \mathbb{I}%
_{P}\otimes \left( \bigotimes_{\substack{ i=1  \\ i\neq j}}^{N}\mathbb{I}%
_{i}\right) \right)  \label{4-1.1}
\end{equation}%
where (see eq. (\ref{3.7.3})) 
\begin{equation}
O_{P_{j}}=\epsilon _{\uparrow \uparrow }^{(j)}\,|\uparrow _{j}\rangle
\langle \uparrow _{j}|+\epsilon _{\downarrow \downarrow }^{(j)}\,|\downarrow
_{j}\rangle \langle \downarrow _{j}|+\epsilon _{\downarrow \uparrow
}^{(j)}\,|\downarrow _{j}\rangle \langle \uparrow _{j}|+\epsilon _{\uparrow
\downarrow }^{(j)}\,|\uparrow _{j}\rangle \langle \downarrow _{j}|
\label{4-1.2}
\end{equation}%
$\mathbb{I}_{P}$ is the identity operator on the subspace $\mathcal{H}_{P}$,
and the coefficients $\epsilon _{\uparrow \uparrow }^{(j)}$, $\epsilon
_{\downarrow \downarrow }^{(j)}$, $\epsilon _{\downarrow \uparrow }^{(j)}$
are now generic. The expectation value of the observables $O_{R}$ in the
state $\left\vert \psi (t)\right\rangle $ is given by%
\begin{equation}
\langle O_{R}\rangle _{\psi (t)}=\langle \psi (t)|O_{R_{j}}|\psi (t)\rangle
=\left\vert \alpha _{j}\right\vert ^{2}\epsilon _{\uparrow \uparrow
}^{(j)}+\left\vert \beta _{j}\right\vert ^{2}\epsilon _{\downarrow
\downarrow }^{(j)}+2\func{Re}\left( \alpha _{j}\beta _{j}^{\ast }\epsilon
_{\downarrow \uparrow }^{(j)}e^{ig_{j}t}\right)  \label{4-1.3}
\end{equation}%
Here there is no need of numerical simulations to see that the third term of
eq. (\ref{4-1.3}) is an oscillating function which, as a consequence, has no
limit for $t\rightarrow \infty $. This result is not surprising since, in
this case, the particle $P_{j}$ is uncoupled to the particles of its
environment.

\section{A generalized spin-bath model: presentation of the model}

Let us consider a closed system $U=A\cup B$ where:

\begin{enumerate}
\item[(i)] The subsystem $A$ is composed of $M$ spin-1/2 particles $A_{i}$,
with $i=1,2,...,M$, each one of them represented in its Hilbert space $%
\mathcal{H}_{A_{i}}$. In each $A_{i}$, the two eigenstates of the spin
operator $S_{A_{i},\overrightarrow{v}}$ in direction $\overrightarrow{v}$\
are $\left\vert \Uparrow _{i}\right\rangle $ and $\left\vert \Downarrow
_{i}\right\rangle $:%
\begin{equation}
S_{A_{i},\overrightarrow{v}}\left\vert \Uparrow _{i}\right\rangle =\frac{1}{2%
}\left\vert \Uparrow _{i}\right\rangle \text{ \ \ \ \ \ \ \ }S_{A_{i},%
\overrightarrow{v}}\left\vert \Downarrow _{i}\right\rangle =-\frac{1}{2}%
\left\vert \Downarrow _{i}\right\rangle  \label{4.1}
\end{equation}%
The Hilbert space of $A$ is $\mathcal{H}_{A}=\bigotimes\limits_{i=1}^{M}%
\mathcal{H}_{A_{i}}$. Then, a pure initial state of $A$ reads%
\begin{equation}
\left\vert \psi _{A}\right\rangle =\bigotimes_{i=1}^{M}\left(
a_{i}\left\vert \Uparrow _{i}\right\rangle +b_{i}\left\vert \Downarrow
_{i}\right\rangle \right) ,\ \ \text{with\ }\left\vert a_{i}\right\vert
^{2}+\left\vert b_{i}\right\vert ^{2}=1  \label{4.2}
\end{equation}

\item[(ii)] The subsystem $B$ is composed of $N$ spin-1/2 particles $B_{k}$,
with $k=1,2,...,N$, each one of them represented in its Hilbert space $%
\mathcal{H}_{B_{k}}$. In each $B_{k}$, the two eigenstates of the spin
operator $S_{B_{k},\overrightarrow{v}}$ in direction $\overrightarrow{v}$\
are $\left\vert \uparrow _{k}\right\rangle $ and $\left\vert \downarrow
_{k}\right\rangle $:%
\begin{equation}
S_{B_{k},\overrightarrow{v}}\left\vert \uparrow _{k}\right\rangle =\frac{1}{2%
}\left\vert \uparrow _{k}\right\rangle \text{ \ \ \ \ \ \ \ }S_{B_{k},%
\overrightarrow{v}}\left\vert \downarrow _{k}\right\rangle =-\frac{1}{2}%
\left\vert \downarrow _{k}\right\rangle  \label{4.3}
\end{equation}%
The Hilbert space of $B$ is $\mathcal{H}_{B}=\bigotimes\limits_{k=1}^{N}%
\mathcal{H}_{B_{k}}$. Then, a pure initial state of $B$ reads%
\begin{equation}
\left\vert \psi _{B}\right\rangle =\bigotimes_{k=1}^{N}\left( \alpha
_{k}\left\vert \uparrow _{k}\right\rangle +\beta _{k}\left\vert \downarrow
_{k}\right\rangle \right) \text{, \ \ with \ \ }\left\vert \alpha
_{k}\right\vert ^{2}+\left\vert \beta _{k}\right\vert ^{2}=1  \label{4.4}
\end{equation}
\end{enumerate}

The Hilbert space of the composite system $U=A\cup B$\ is, then, 
\begin{equation}
\mathcal{H}=\mathcal{H}_{A}\otimes \mathcal{H}_{B}=\left(
\bigotimes\limits_{i=1}^{M}\mathcal{H}_{A_{i}}\right) \otimes \left(
\bigotimes\limits_{k=1}^{N}\mathcal{H}_{B_{k}}\right)  \label{4.5}
\end{equation}%
Therefore, from eqs. (\ref{4.2}) and (\ref{4.4}), a pure initial state of $U$
reads%
\begin{equation}
\left\vert \psi _{0}\right\rangle =\left\vert \psi _{A}\right\rangle \otimes
\left\vert \psi _{B}\right\rangle =\left( \bigotimes_{i=1}^{M}\left(
a_{i}\left\vert \Uparrow _{i}\right\rangle +b_{i}\left\vert \Downarrow
_{i}\right\rangle \right) \right) \otimes \left( \bigotimes_{k=1}^{N}\left(
\alpha _{k}\left\vert \uparrow _{k}\right\rangle +\beta _{k}\left\vert
\downarrow _{k}\right\rangle \right) \right)  \label{4.6}
\end{equation}

As in the original spin-bath model, the self-Hamiltonians $H_{A_{i}}$ and $%
H_{B_{k}}$ are taken to be zero. In turn, there is no interaction among the
particles $A_{i}$ nor among the particles $B_{k}$. As a consequence, the
total Hamiltonian $H$ of the composite system $U$ is given by%
\begin{equation}
H=H_{A}\otimes H_{B}=\left( \sum_{i=1}^{M}\left[ \frac{1}{2}\left(
\left\vert \Uparrow _{i}\right\rangle \left\langle \Uparrow _{i}\right\vert
-\left\vert \Downarrow _{i}\right\rangle \left\langle \Downarrow
_{i}\right\vert \right) \otimes \left( \bigotimes_{j\neq i}^{M}\mathbb{I}%
_{A_{j}}\right) \right] \right) \otimes \left( \sum_{k=1}^{N}\left[
g_{k}\left( \left\vert \uparrow _{k}\right\rangle \left\langle \uparrow
_{k}\right\vert -\left\vert \downarrow _{k}\right\rangle \left\langle
\downarrow _{k}\right\vert \right) \otimes \left( \bigotimes_{l\neq k}^{N}%
\mathbb{I}_{B_{l}}\right) \right] \right)  \label{4.7}
\end{equation}%
where $\mathbb{I}_{A_{j}}=\left\vert \Uparrow _{j}\right\rangle \left\langle
\Uparrow _{j}\right\vert +\left\vert \Downarrow _{j}\right\rangle
\left\langle \Downarrow _{j}\right\vert $ is the identity on the subspace $%
\mathcal{H}_{A_{j}}$ and$\ \mathbb{I}_{B_{l}}=\left\vert \uparrow
_{l}\right\rangle \left\langle \uparrow _{l}\right\vert +\left\vert
\downarrow _{l}\right\rangle \left\langle \downarrow _{l}\right\vert $ is
the identity on the subspace $\mathcal{H}_{B_{l}}$. Let us notice that the
eq. (\ref{3.4}) of the original model is the particular case of eq. (\ref%
{4.7}) for $M=1$. This Hamiltonian describes a situation where the particles
of $A$ do not interact to each other, the same holds for the particles of $B$%
, but each particle of $A$ interacts with all the particles of $B$ and vice
versa, as shown in Figure 1.%

\begin{figure}[t]
 \centerline{\scalebox{0.7}{\includegraphics{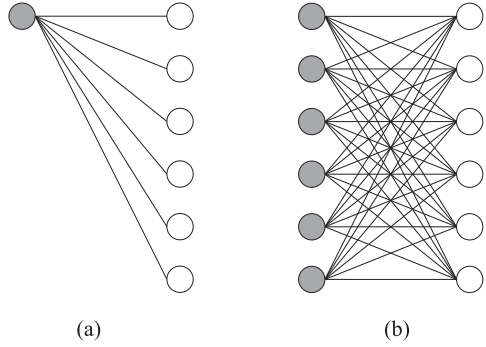}}}
\caption{Figure 1: Schema of the
interactions among the particles of the open system $A$ (grey circles) and
of the open system $B$ (white circles): (a) original spin-bath model ($M=1$%
), and (b) generalized spin-bath model ($M\neq 1$).}
 \label{fig 1}\vspace*{0.cm}
\end{figure}

In eq. (\ref{4.7}), $H$ is written in its diagonal form; then, the energy
eigenvectors are%
\begin{align}
& \left\vert \Uparrow _{1}\right\rangle ...\left\vert \Uparrow
_{i}\right\rangle ...\left\vert \Uparrow _{M-1}\right\rangle \left\vert
\Uparrow _{M}\right\rangle \left\vert \uparrow _{1}\right\rangle
...\left\vert \uparrow _{k}\right\rangle ...\left\vert \uparrow
_{N-1}\right\rangle \left\vert \uparrow _{N}\right\rangle  \notag \\
& \left\vert \Uparrow _{1}\right\rangle ...\left\vert \Uparrow
_{i}\right\rangle ...\left\vert \Uparrow _{M-1}\right\rangle \left\vert
\Uparrow _{M}\right\rangle \left\vert \uparrow _{1}\right\rangle
...\left\vert \uparrow _{k}\right\rangle ...\left\vert \uparrow
_{N-1}\right\rangle \left\vert \downarrow _{N}\right\rangle  \notag \\
& ...  \notag \\
& \left\vert \Downarrow _{1}\right\rangle ...\left\vert \Downarrow
_{i}\right\rangle ...\left\vert \Downarrow _{M-1}\right\rangle \left\vert
\Downarrow _{M}\right\rangle \left\vert \downarrow _{1}\right\rangle
...\left\vert \downarrow _{k}\right\rangle ...\left\vert \downarrow
_{N-1}\right\rangle \left\vert \downarrow _{N}\right\rangle  \label{4.8}
\end{align}%
In turn, the eigenvectors of $H_{A}$ form a basis of $\mathcal{H}_{A}$. In
order to simplify the expressions, we will introduce a particular
arrangement into the set of those vectors, by calling them $\left\vert 
\mathcal{A}_{i}\right\rangle $:\ the set $\left\{ \left\vert \mathcal{A}%
_{i}\right\rangle \right\} $ is an eigenbasis of $H_{A}$ with $2^{M}$
elements. The $\left\vert \mathcal{A}_{i}\right\rangle $ will be ordered in
terms of the number $l\in 
\mathbb{N}
_{0}$ of particles of $A$ having spin $\left\vert \Downarrow \right\rangle $%
. Then, we have that:

\begin{itemize}
\item $l=0$ corresponds to the unique state with all the particles with spin 
$\left\vert \Uparrow \right\rangle $:%
\begin{equation}
\left\vert \mathcal{A}_{1}\right\rangle =\left\vert \Uparrow ,\Uparrow
,...,\Uparrow ,\Uparrow \right\rangle \Longrightarrow H_{A}\left\vert 
\mathcal{A}_{1}\right\rangle =\frac{M}{2}\left\vert \mathcal{A}%
_{1}\right\rangle  \label{4.9}
\end{equation}

\item $l=1$ corresponds to the $M$ states with only one particle with spin $%
\left\vert \Downarrow \right\rangle $. Since the order of the eigenvectors
with the same eigenvalue will be irrelevant for the computations, we will
order these states in an arbitrary way:%
\begin{eqnarray}
\left\vert \mathcal{A}_{j}\right\rangle &=&\left\vert \Uparrow ,\Uparrow
,...,\Uparrow ,\Downarrow ,\Uparrow ,...,\Uparrow ,\Uparrow \right\rangle
\Longrightarrow H_{A}\left\vert \mathcal{A}_{j}\right\rangle =\frac{M-2}{2}%
\left\vert \mathcal{A}_{j}\right\rangle  \notag \\
\text{with \ \ }j &=&2,3,...,M+1  \label{4.10}
\end{eqnarray}

\item $l=2$ corresponds to the $\frac{\left( M-1\right) M}{2}$ states with
two particles with spin $\left\vert \Downarrow \right\rangle $. Again, we
will order these states in an arbitrary way:%
\begin{align}
\left\vert \mathcal{A}_{j}\right\rangle & =\left\vert \Uparrow ,\Uparrow
,,...,\Uparrow ,\Downarrow ,\Uparrow ,,...,\Uparrow ,\Downarrow ,\Uparrow
,...,\Uparrow ,\Uparrow \right\rangle \Longrightarrow H_{A}\left\vert 
\mathcal{A}_{j}\right\rangle =\frac{M-4}{2}\left\vert \mathcal{A}%
_{j}\right\rangle  \notag \\
\text{with \ \ }j& =M+2,M+3,...,M+1+\frac{\left( M-1\right) M}{2}
\label{4.11}
\end{align}

\item For the remaining values of $l$, the procedure is analogous.\medskip
\end{itemize}

Consequently, we have:%
\begin{align}
& 1\text{ eigenvector with eigenvalue }\frac{M}{2}  \notag \\
& M\text{ eigenvectors with eigenvalue }\frac{M-2}{2}  \notag \\
& \vdots  \notag \\
& \frac{M!}{(M-l)!l!}\text{ eigenvectors with eigenvalue }\frac{M-2l}{2}
\label{4.12}
\end{align}%
with $l=0,1,...M$. Then, it is clear that $H_{A}$ is degenerate: it has $%
2^{M}$ eigenvectors but only $M$ different eigenvalues. Therefore, a generic
state $\left\vert \mathcal{A}\right\rangle $ of the system $A$ can be
written in the basis $\left\{ \left\vert \mathcal{A}_{i}\right\rangle
\right\} $ as%
\begin{equation}
\left\vert \mathcal{A}\right\rangle =\sum_{i=1}^{2^{M}}C_{i}\left\vert 
\mathcal{A}_{i}\right\rangle \in \mathcal{H}_{A}\text{ \ \qquad\ with \ \ }%
\sum_{i=1}^{2^{M}}\left\vert C_{i}\right\vert ^{2}=1  \label{4.13}
\end{equation}%
By introducing eq. (\ref{4.13}) into eq. (\ref{4.6}), a pure initial state
of the composite system $U=A\cup B$ reads%
\begin{equation}
\left\vert \psi _{0}\right\rangle =\left( \sum_{i=1}^{2^{M}}C_{i}\left\vert 
\mathcal{A}_{i}\right\rangle \right) \otimes \left(
\bigotimes_{k=1}^{N}\left( \alpha _{k}\left\vert \uparrow _{k}\right\rangle
+\beta _{k}\left\vert \downarrow _{k}\right\rangle \right) \right)
\label{4.14}
\end{equation}%
If we group the degrees of freedom of $B$ in a single ket $\left\vert 
\mathcal{B}(0)\right\rangle $, $\left\vert \psi _{0}\right\rangle $ results%
\begin{equation}
|\psi _{0}\rangle =\sum_{i=1}^{2^{M}}C_{i}\left\vert \mathcal{A}%
_{i}\right\rangle \otimes \left\vert \mathcal{B}(0)\right\rangle
\label{4.15}
\end{equation}%
The time-evolution of $|\psi (t)\rangle $ is ruled by the time-evolution
operator $\mathcal{U}(t)=e^{-iHt}=e^{-i(H_{A}\otimes H_{B})t}$:%
\begin{equation}
|\psi (t)\rangle =\mathcal{U}(t)|\psi _{0}\rangle
=\sum_{i=1}^{2^{M}}C_{i}\,e^{-i(H_{A}\otimes H_{B})t}\,\left\vert \mathcal{A}%
_{i}\right\rangle \otimes \left\vert \mathcal{B}(0)\right\rangle
=\sum_{i=1}^{2^{M}}C_{i}\,e^{-iH_{A}t}\,\left\vert \mathcal{A}%
_{i}\right\rangle \otimes e^{-iH_{B}t}\,\left\vert \mathcal{B}%
(0)\right\rangle  \label{4.15-1}
\end{equation}%
If we use $\Lambda _{k}$ to denote the eigenvalue of $H_{A}$ corresponding
to the eigenvector $\left\vert \mathcal{A}_{k}\right\rangle $, then%
\begin{equation}
|\psi (t)\rangle =\sum_{i=1}^{2^{M}}C_{i}\,\left\vert \mathcal{A}%
_{i}\right\rangle \otimes e^{-i\Lambda _{i}H_{B}t}\,\left\vert \mathcal{B}%
(0)\right\rangle =\sum_{i=1}^{2^{M}}C_{i}\,\left\vert \mathcal{A}%
_{i}\right\rangle \otimes \left\vert \mathcal{B}(t)\right\rangle
\label{4.15-2}
\end{equation}%
where (see eq. (\ref{4.7}))%
\begin{equation}
\left\vert \mathcal{B}(t)\right\rangle =e^{-i\Lambda _{i}H_{B}t}\,\left\vert 
\mathcal{B}(0)\right\rangle =\exp \left[ -i\Lambda
_{k}\sum\limits_{j=1}^{N}g_{j}\left( \left\vert \uparrow _{j}\right\rangle
\left\langle \uparrow _{j}\right\vert -\left\vert \downarrow
_{j}\right\rangle \left\langle \downarrow _{j}\right\vert \right) t\right]
\left\vert \mathcal{B}(0)\right\rangle  \label{4.16}
\end{equation}%
Since the number of the eigenstates of $H_{A}$ with the same eigenvalue is
given by eqs. (\ref{4.12}), the terms of $|\psi (t)\rangle $ can be arranged
as%
\begin{align}
\left\vert \psi (t)\right\rangle & =\left( C_{1}\left\vert \mathcal{A}%
_{1}\right\rangle \left\vert \mathcal{B}_{0}(t)\right\rangle \right) +\left(
\sum\limits_{\lambda =1}^{M+1}C_{\lambda }\left\vert \mathcal{A}_{\lambda
}\right\rangle \left\vert \mathcal{B}_{1}(t)\right\rangle \right) +\left(
\sum\limits_{\lambda =M+2}^{M+1+\frac{\left( M-1\right) M}{2}}C_{\lambda
}\left\vert \mathcal{A}_{\lambda }\right\rangle \left\vert \mathcal{B}%
_{2}(t)\right\rangle \right) +...+  \notag \\
& +\left( \sum\limits_{\lambda =1+\sum_{p=0}^{l-1}\binom{M}{P}%
}^{\sum_{p=0}^{l}\binom{M}{P}}C_{\lambda }\left\vert \mathcal{A}_{\lambda
}\right\rangle \left\vert \mathcal{B}_{l}(t)\right\rangle \right)
+...+\left( C_{2^{M}}\left\vert \mathcal{A}_{2^{M}}\right\rangle \left\vert 
\mathcal{B}_{M}(t)\right\rangle \right)  \label{4.17}
\end{align}%
where%
\begin{equation}
\left\vert \mathcal{B}_{l}(t)\right\rangle
=\bigotimes\limits_{k=1}^{N}\left( \alpha _{k}e^{i\frac{\left( 2l-M\right) }{%
2}g_{k}t}\left\vert \uparrow _{k}\right\rangle +\beta _{k}e^{-i\frac{\left(
2l-M\right) }{2}g_{k}t}\left\vert \downarrow _{k}\right\rangle \right)
\label{4.18}
\end{equation}%
If we compare eq. (\ref{4.18}) with eq. (\ref{3.6}), we can see that $%
\left\vert \mathcal{E}_{\Uparrow }(t)\right\rangle $ and $\left\vert 
\mathcal{E}_{\Downarrow }(t)\right\rangle $ are the particular cases of $%
\left\vert \mathcal{B}_{l}(t)\right\rangle $ for $M=1$ and, then, $l=0,1$.
Let us recall that $l$ is the number of particles of the system $A$ having
spin $\left\vert \Downarrow \right\rangle $. Then, with $M=1$ and $l=0$, $%
\left\vert \mathcal{B}_{l}(t)\right\rangle =\left\vert \mathcal{E}_{\Uparrow
}(t)\right\rangle $, and with $M=1$ and $l=1$, $\left\vert \mathcal{B}%
_{l}(t)\right\rangle =\left\vert \mathcal{E}_{\Downarrow }(t)\right\rangle $.

If we define the function%
\begin{equation}
f(l)=\QATOPD\{ \} {\sum_{p=0}^{l}\binom{M}{P}\text{ if }l=0,1,...,M}{0\text{
\ \ \ \ \ \ \ \ \ \ otherwise}}  \label{4-19}
\end{equation}%
then eq. (\ref{4.17}) can be rewritten as%
\begin{equation}
\left\vert \psi (t)\right\rangle =\sum\limits_{l=0}^{M}\sum\limits_{\lambda
=f(l-1)+1}^{f(l)}C_{\lambda }\left\vert \mathcal{A}_{\lambda }\right\rangle
\left\vert \mathcal{B}_{l}(t)\right\rangle  \label{4.20}
\end{equation}%
and the state operator $\rho (t)=\left\vert \psi (t)\right\rangle
\left\langle \psi (t)\right\vert $ reads%
\begin{equation}
\rho (t)=\sum\limits_{l,l^{\prime }=0}^{M}\sum\limits_{\QATOP{\lambda
=f(l-1)+1}{\lambda ^{\prime }=f(l^{\prime }-1)+1}}^{\QATOP{f(l)}{f(l^{\prime
})}}C_{\lambda }C_{\lambda ^{\prime }}^{\ast }\left\vert \mathcal{A}%
_{\lambda }\right\rangle \left\vert \mathcal{B}_{l}(t)\right\rangle
\left\langle \mathcal{B}_{l^{\prime }}(t)\right\vert \left\langle \mathcal{A}%
_{\lambda ^{\prime }}\right\vert  \label{4.21}
\end{equation}

An observable $O\in \mathcal{O}=\mathcal{H}\otimes \mathcal{H}$ of the
closed system $U=A\cup B$ can be expressed as%
\begin{equation}
O=\left( \sum\limits_{\lambda ,\lambda ^{\prime }=0}^{2^{M}}s_{\lambda
,\lambda ^{\prime }}\left\vert \mathcal{A}_{\lambda }\right\rangle
\left\langle \mathcal{A}_{\lambda ^{\prime }}\right\vert \right) \otimes
\left( \bigotimes_{i=1}^{N}\left( \epsilon _{\uparrow \uparrow
}^{(i)}\left\vert \uparrow _{i}\right\rangle \left\langle \uparrow
_{i}\right\vert +\epsilon _{\uparrow \downarrow }^{(i)}\left\vert \uparrow
_{i}\right\rangle \left\langle \downarrow _{i}\right\vert +\epsilon
_{\downarrow \uparrow }^{(i)}\left\vert \downarrow _{i}\right\rangle
\left\langle \uparrow _{i}\right\vert +\epsilon _{\downarrow \downarrow
}^{(i)}\left\vert \downarrow _{i}\right\rangle \left\langle \downarrow
_{i}\right\vert \right) \right)  \label{4.22}
\end{equation}%
Let us notice that eq. (\ref{3.7.1}) (a generic observable in the original
spin-bath model) is a particular case of this eq. (\ref{4.22}), with only
four terms in the first factor. Analogously to that case, the diagonal
components $s_{\lambda ,\lambda }$, $\epsilon _{\uparrow \uparrow }^{(i)}$, $%
\epsilon _{\downarrow \downarrow }^{(i)}$ are real numbers, and the
off-diagonal components are complex numbers satisfying $s_{\lambda ,\lambda
^{\prime }}=s_{\lambda ^{\prime },\lambda }^{\ast }$, $\epsilon _{\uparrow
\downarrow }^{(i)}=\epsilon _{\downarrow \uparrow }^{(i)\ast }$. Then, the
expectation value of the observable $O$ in the state $\rho (t)$ of eq. (\ref%
{4.21}) can be computed as%
\begin{equation}
\langle O\rangle _{\rho (t)}=Tr\left( O\rho (t)\right)
=\sum\limits_{l,l^{\prime }=0}^{M}\sum\limits_{\QATOP{\lambda =f(l-1)+1}{%
\lambda ^{\prime }=f(l^{\prime }-1)+1}}^{\QATOP{f(l)}{f(l^{\prime })}%
}B_{\lambda ,\lambda ^{\prime }}T_{l,l^{\prime }}(t)  \label{4.23}
\end{equation}%
where%
\begin{equation}
T_{l,l^{\prime }}(t)=\prod_{j=1}^{N}\left[ \left\vert \alpha _{j}\right\vert
^{2}\epsilon _{\uparrow \uparrow }^{(j)}e^{i\left( g_{j,l}-g_{j,l^{\prime
}}\right) \frac{t}{2}}+\left\vert \beta _{j}\right\vert ^{2}\epsilon
_{\downarrow \downarrow }^{(j)}e^{-i\left( g_{j,l}-g_{j,l^{\prime }}\right) 
\frac{t}{2}}+2\func{Re}\left( \alpha _{j}\beta _{j}^{\ast }\epsilon
_{\downarrow \uparrow }^{(j)}e^{i\left( g_{j,l}+g_{j,l^{\prime }}\right) 
\frac{t}{2}}\right) \right]  \label{4.24}
\end{equation}%
and 
\begin{equation}
g_{j,l}=\left( 2l-M\right) g_{j}\text{, \ \qquad\ }B_{\lambda ,\lambda
^{\prime }}=C_{\lambda }C_{\lambda ^{\prime }}^{\ast }s_{\lambda ^{\prime
},\lambda }  \label{4.25}
\end{equation}%
Since the exponents in eq. (\ref{4.24}) are of the form $g_{j,l}\pm
g_{j,l^{\prime }}$, in some cases they are zero. So, we can write%
\begin{equation}
\langle O\rangle _{\rho (t)}=\sum\limits_{l=0}^{M}\sum\limits_{\QATOP{%
\lambda =f(l-1)+1}{\lambda ^{\prime }=f(l-1)+1}}^{f(l)}B_{\lambda ,\lambda
^{\prime }}T_{l,l}(t)+\sum\limits_{l=0}^{\tilde{M}}\sum\limits_{\QATOP{%
\lambda =f(l-1)+1}{\lambda ^{\prime }=f(M-l-1)+1}}^{\QATOP{f(l)}{f(M-l)}%
}B_{\lambda ,\lambda ^{\prime }}2\func{Re}\left( T_{l,M-l}(t)\right)
+\sum\limits_{\QATOP{\QATOP{l,l^{\prime }=0}{l\neq l^{\prime }}}{l^{\prime
}\neq M-l}}^{M}\sum\limits_{\QATOP{\lambda =f(l-1)+1}{\lambda ^{\prime
}=f(l-1)+1}}^{\QATOP{f(l)}{f(l^{\prime })}}B_{\lambda ,\lambda ^{\prime
}}T_{l,l^{\prime }}(t)  \label{4.26}
\end{equation}%
where%
\begin{equation}
\tilde{M}=\QATOPD\{ \} {\frac{M-2}{2}\text{ if }M\text{ is even}}{\frac{M-1}{%
2}\text{\ if }M\text{ is odd}}  \label{4.27}
\end{equation}%
\begin{equation}
T_{l,l}(t)=\prod_{j=1}^{N}\left[ \left\vert \alpha _{j}\right\vert
^{2}\epsilon _{\uparrow \uparrow }^{(j)}+\left\vert \beta _{j}\right\vert
^{2}\epsilon _{\downarrow \downarrow }^{(j)}+2\func{Re}\left( \alpha
_{j}\beta _{j}^{\ast }\epsilon _{\downarrow \uparrow
}^{(j)}e^{ig_{j,l}t}\right) \right]  \label{4.28}
\end{equation}%
\begin{equation}
T_{l,M-l}(t)=\prod_{j=1}^{N}\left[ \left\vert \alpha _{j}\right\vert
^{2}\epsilon _{\uparrow \uparrow }^{(j)}e^{ig_{j,l}t}+\left\vert \beta
_{j}\right\vert ^{2}\epsilon _{\downarrow \downarrow }^{(j)}e^{-ig_{j,l}t}+2%
\func{Re}\left( \alpha _{j}\beta _{j}^{\ast }\epsilon _{\downarrow \uparrow
}^{(j)}\right) \right]  \label{4.29}
\end{equation}%
Let us notice that eqs. (\ref{4.28}) and (\ref{4.29}) are analogous to eqs. (%
\ref{3.9}) and (\ref{3.10}) for $\Gamma _{0}(t)$ and $\Gamma _{1}(t)$,
respectively, in the original model, with $g_{j,l}=\left( 2l-M\right) g_{j}$
instead of $g_{j}$. In particular, when $M=1$ and, so, $l=0,1$, then $%
T_{l,l}(t)=\Gamma _{0}(t)$ and $T_{l,M-l}(t)=\Gamma _{1}(t)$.

As in the case of the original spin-bath model, here we will consider
different meaningful ways of selecting the relevant observables.

\section{Generalized spin-bath model: Decomposition 1}

\subsection{Selecting the relevant observables}

In this case $A$ is the open system $S$ and $B$ is the environment $E$. This
is a generalization of Decomposition 1 in the original spin-bath model. The
only difference with respect to that case is that here the system $S$ is
composed of $M\geq 1$ particles instead of only one. Then, the decomposition
for this case is%
\begin{equation}
\mathcal{H}=\mathcal{H}_{S}\otimes \mathcal{H}_{E}=\left(
\bigotimes\limits_{i=1}^{M}\mathcal{H}_{A_{i}}\right) \otimes \left(
\bigotimes\limits_{k=1}^{N}\mathcal{H}_{B_{k}}\right)  \label{5.0}
\end{equation}%
Therefore, the relevant observables $O_{R}$ of the closed system $U$ are
those corresponding to $A$, and they are obtained from eq. (\ref{4.22}) by
making $\epsilon _{\uparrow \uparrow }^{(i)}=\epsilon _{\downarrow
\downarrow }^{(i)}=1,$ $\epsilon _{\uparrow \downarrow }^{(i)}=0$ (compare
with eq. (\ref{3.11}) in the original spin-bath model):%
\begin{equation}
O_{R}=O_{S}\otimes \mathbb{I}_{E}=\left( \sum\limits_{\lambda ,\lambda
^{\prime }=0}^{2^{M}}s_{\lambda ,\lambda ^{\prime }}\left\vert \mathcal{A}%
_{\lambda }\right\rangle \left\langle \mathcal{A}_{\lambda ^{\prime
}}\right\vert \right) \otimes \left( \bigotimes_{i=1}^{N}\mathbb{I}%
_{i}\right)  \label{5.1}
\end{equation}%
With this condition, the expectation values of these observables are given
by eq. (\ref{4.26}), with%
\begin{eqnarray}
T_{l,l}(t) &=&\prod_{j=1}^{N}\left( \left\vert \alpha _{j}\right\vert
^{2}+\left\vert \beta _{j}\right\vert ^{2}\right) =1  \label{5.2} \\
T_{l,M-l}(t) &=&\prod_{j=1}^{N}\left( \left\vert \alpha _{j}\right\vert
^{2}e^{ig_{j,l}t}+\left\vert \beta _{j}\right\vert ^{2}e^{-ig_{j,l}t}\right)
\label{5.3} \\
T_{l,l^{\prime }}(t) &=&\prod_{j=1}^{N}\left( \left\vert \alpha
_{j}\right\vert ^{2}e^{i\left( g_{j,l}-g_{j,l^{\prime }}\right) \frac{t}{2}%
}+\left\vert \beta _{j}\right\vert ^{2}e^{-i\left( g_{j,l}-g_{j,l^{\prime
}}\right) \frac{t}{2}}\right)  \label{5.4}
\end{eqnarray}%
If we define the functions $R_{l}(t)=|T_{l,M-l}(t)|^{2}$ and $R_{ll^{\prime
}}(t)=|T_{l,l^{\prime }}(t)|^{2}$, they result%
\begin{eqnarray}
R_{l}(t) &=&\prod_{j=1}^{N}\left( \left\vert \alpha _{j}\right\vert
^{4}+\left\vert \beta _{j}\right\vert ^{4}+2\left\vert \alpha
_{j}\right\vert ^{2}\left\vert \beta _{j}\right\vert ^{2}\cos \left( 2\left(
2l-M\right) g_{j}t\right) \right)  \label{5.5} \\
R_{ll^{\prime }}(t) &=&\prod_{j=1}^{N}\left( \left\vert \alpha
_{j}\right\vert ^{4}+\left\vert \beta _{j}\right\vert ^{4}+2\left\vert
\alpha _{j}\right\vert ^{2}\left\vert \beta _{j}\right\vert ^{2}\cos \left(
2\left( l-l^{\prime }\right) g_{j}t\right) \right)  \label{5.6}
\end{eqnarray}%
We can see that $|r(t)|^{2}$ of eq. (\ref{3.14}) in the original model is
the particular case of $R_{l}(t)$ for $M=1$.

\subsection{Computing the behavior of the relevant expectation values}

The expectation value given by eq. (\ref{4.26}) has three terms, $\langle
O_{R}\rangle _{\rho (t)}=\Sigma ^{\left( 1\right) }+\Sigma ^{\left( 2\right)
}+\Sigma ^{\left( 3\right) }$, which can be analyzed separately:

\begin{itemize}
\item From eq. (\ref{5.2}), the first term reads%
\begin{equation}
\Sigma ^{\left( 1\right) }=\sum\limits_{l=0}^{M}\sum\limits_{\lambda
,\lambda ^{\prime }=f(l-1)+1}^{f(l)}B_{\lambda ,\lambda ^{\prime
}}=\sum\limits_{l=0}^{M}\sum\limits_{\lambda ,\lambda ^{\prime
}=f(l-1)+1}^{f(l)}C_{\lambda }C_{\lambda ^{\prime }}^{\ast }s_{\lambda
^{\prime },\lambda }\neq \Sigma ^{\left( 1\right) }(t)  \label{5.7}
\end{equation}%
It is clear that this first term does not evolve with time.

\item The time-dependence of the second term is given by $T_{l,M-l}(t)$:%
\begin{equation}
\Sigma ^{\left( 2\right) }(t)=\sum\limits_{l=0}^{\tilde{M}}\sum\limits_{%
\QATOP{\lambda =f(l-1)+1}{\lambda ^{\prime }=f(M-l-1)+1}}^{\QATOP{f(l)}{%
f(M-l)}}B_{\lambda ,\lambda ^{\prime }}2\func{Re}\left( T_{l,M-l}(t)\right)
\label{5.8}
\end{equation}%
Then, in order to obtain the limit of this term, we have to compute the
limit of $R_{l}(t)=|T_{l,M-l}(t)|^{2}$ of eq. (\ref{5.5}). As in the case of
the original spin-bath model, here we take $\left\vert \alpha
_{j}\right\vert ^{2}$ and $\left\vert \beta _{ji}\right\vert ^{2}$ as random
numbers in the closed interval $\left[ 0,1\right] $, such that $|\alpha
_{j}|^{2}+|\beta _{j}|^{2}=1$. Then%
\begin{eqnarray}
\max_{t}\left( \left\vert \alpha _{j}\right\vert ^{4}+\left\vert \beta
_{j}\right\vert ^{4}+2\left\vert \alpha _{j}\right\vert ^{2}\left\vert \beta
_{j}\right\vert ^{2}\cos \left( 2\left( 2l-M\right) g_{j}t\right) \right)
&=&1  \label{5.9} \\
\min_{t}\left( \left\vert \alpha _{j}\right\vert ^{4}+\left\vert \beta
_{j}\right\vert ^{4}+2\left\vert \alpha _{j}\right\vert ^{2}\left\vert \beta
_{j}\right\vert ^{2}\cos \left( 2\left( 2l-M\right) g_{j}t\right) \right)
&=&\left( 2\left\vert \alpha _{j}\right\vert ^{2}-1\right) ^{2}  \label{5.10}
\end{eqnarray}%
Therefore, $\left[ \left\vert \alpha _{j}\right\vert ^{4}+\left\vert \beta
_{j}\right\vert ^{4}+2\left\vert \alpha _{j}\right\vert ^{2}\left\vert \beta
_{j}\right\vert ^{2}\cos \left( 2\left( 2l-M\right) g_{j}t\right) \right] $
is a random number which, if $t\neq 0$, fluctuates between $1$ and $\left(
2\left\vert \alpha _{j}\right\vert ^{2}-1\right) ^{2}$. Again, when the
environment has many particles (that is, when $N\rightarrow \infty $), the
statistical value of the cases $\left\vert \alpha _{j}\right\vert ^{2}=1$, $%
\left\vert \beta _{j}\right\vert ^{2}=1$, $\left\vert \alpha _{j}\right\vert
^{2}=0$ and $\left\vert \beta _{j}\right\vert ^{2}=0$ tends to zero. In this
situation, eq. (\ref{5.5}) for $R_{l}(t)$ is an infinite product of numbers
belonging to the open interval $\left( 0,1\right) $. \ As a consequence,
when $N\rightarrow \infty $, $R_{l}(t)\rightarrow 0$.

\item The time-dependence of the third term is given by $T_{l,l^{\prime
}}(t) $:%
\begin{equation}
\Sigma ^{\left( 3\right) }(t)=\sum\limits_{\QATOP{\QATOP{l,l^{\prime }=0}{%
l\neq l^{\prime }}}{l^{\prime }\neq M-l}}^{M}\sum\limits_{\QATOP{\lambda
=f(l-1)+1}{\lambda ^{\prime }=f(l^{\prime }-1)+1}}^{\QATOP{f(l)}{f(l^{\prime
})}}B_{\lambda ,\lambda ^{\prime }}T_{l,l^{\prime }}(t)  \label{5.11}
\end{equation}%
with the restrictions on $l$ and $l^{\prime }$: $\ l\neq l^{\prime }$ and $%
l^{\prime }\neq M-l$. As in the second term, we have to compute the limit of 
$R_{ll^{\prime }}(t)=|T_{l,l^{\prime }}(t)|^{2}$ of eq. (\ref{5.6}) and, on
the basis of an analogous argument, the result is the same as above: when $%
N\rightarrow \infty $, $R_{ll^{\prime }}(t)\rightarrow 0$.
\end{itemize}

If we want now to evaluate the limit of $\langle O_{R}\rangle _{\rho (t)}$
for $t\rightarrow \infty $, we have to compute the limits of the second and
the third terms (since the first term, as we have seen, is
time-independent). Here we have to distinguish three cases: $M\ll N$, $M\gg
N $ and $M\simeq N$.\bigskip

\textbf{Case (a): }$M\ll N$

This case is similar to Decomposition 1 in the original spin-bath model,
since in both cases $M\ll N$: the only difference is that in the original
model $M=1$ whereas here $M\geq 1$.

In fact, we have seen that $T_{l,M-l}(t)$ is analogous to $\Gamma _{1}(t)$
in the original model. Moreover, $T_{l,l^{\prime }}(t)$ has the same
functional form as $\Gamma _{1}(t)$. In paper \cite{Max} it is shown that $%
\Gamma _{1}(t)$ approaches zero for $t\rightarrow \infty $. This means that
we can infer that $T_{l,M-l}(t)$ and $T_{l,l^{\prime }}(t)$ also approach
zero for $t\rightarrow \infty $. On the other hand, the terms $\Sigma
^{\left( 2\right) }(t)$ and $\Sigma ^{\left( 3\right) }(t)$ are sums of less
than $M$ terms involving $T_{l,M-l}(t)$ and $T_{l,l^{\prime }}(t)$. As a
consequence, since in this case $M$ is a small number, the sum of a small
number of terms approaching zero for $t\rightarrow \infty $ also approaches
zero: $\lim_{t\rightarrow \infty }\Sigma ^{\left( 2\right) }(t)=0$ and $%
\lim_{t\rightarrow \infty }\Sigma ^{\left( 3\right) }(t)=0$. Therefore,%
\begin{equation}
\lim_{t\rightarrow \infty }\langle O_{R}\rangle _{\rho
(t)}=\lim_{t\rightarrow \infty }\left[ \Sigma ^{\left( 1\right) }(t)+\Sigma
^{\left( 2\right) }(t)+\Sigma ^{\left( 3\right) }(t)\right] =\Sigma ^{\left(
1\right) }(t)  \label{5.11-1}
\end{equation}%
In other words, 
\begin{equation}
\lim_{t\rightarrow \infty }\langle O_{R}\rangle _{\rho
(t)}=\sum\limits_{l=0}^{M}\sum\limits_{\lambda ,\lambda ^{\prime
}=f(l-1)+1}^{f(l)}B_{\lambda ,\lambda ^{\prime
}}=\sum\limits_{l=0}^{M}\sum\limits_{\lambda ,\lambda ^{\prime
}=f(l-1)+1}^{f(l)}C_{\lambda }C_{\lambda ^{\prime }}^{\ast }s_{\lambda
^{\prime },\lambda }=\langle O_{R}\rangle _{\rho _{\ast }}  \label{5.12}
\end{equation}%
where $\rho _{\ast }$ is the final diagonal state of $U$. This result can
also be expressed in terms of the reduced density operator $\rho _{A}$ of
the system $A$ as:%
\begin{equation}
\lim_{t\rightarrow \infty }\langle O_{R}\rangle _{\rho (t)}=\langle
O_{R}\rangle _{\rho _{\ast }}=\lim_{t\rightarrow \infty }\langle
O_{A}\rangle _{\rho _{A}(t)}=\langle O_{A}\rangle _{\rho _{A\ast }}
\label{5.13}
\end{equation}%
In the eigenbasis of the Hamiltonian $H_{A}$ of $A$, the final reduced
density operator $\rho _{A\ast }$ is expressed by a $2^{M}\times 2^{M}$\
matrix:%
\begin{equation}
\rho _{A\ast }=%
\begin{pmatrix}
\rho _{l=0} & 0 & 0 & 0 & ... & 0 \\ 
0 & \rho _{l=1} & 0 & 0 & ... & 0 \\ 
0 & 0 & \rho _{l=2} & 0 & ... & 0 \\ 
0 & 0 & 0 & \rho _{l=3} & ... & 0 \\ 
... & ... & ... & ... & ... & ... \\ 
0 & 0 & 0 & 0 & 0 & \rho _{l=M}%
\end{pmatrix}
\label{5.14}
\end{equation}%
where $\rho _{l=0}=\left\vert C_{1}\right\vert ^{2}$ and each $\rho _{l}$ is
a matrix of dimension $\frac{M!}{\left( M-l\right) !l!}\times \frac{M!}{%
\left( M-l\right) !l!}$.\ This result might seem insufficient for
decoherence because, since the $\rho _{l}$ are matrices, $\rho _{A\ast }$
seems to be non completely diagonal in the eigenbasis of the Hamiltonian $%
H_{A}$. However, we have to recall that all the states $\left\vert \mathcal{A%
}_{i}\right\rangle $ with same $l$ are degenerate eigenvectors corresponding
to the same eigenvalue of $H_{A}$; then, the basis that diagonalizes $\rho
_{A\ast }$ (i.e., that diagonalizes all the matrices $\rho _{l}$) is an
eigenbasis of $H_{A}$. Summing up, the system $S=A$ of $M$ particles in
interaction with its environment $E=B$ of $N\gg M$ particles decoheres in
the eigenbasis of $\rho _{A\ast }$, which is also an eigenbasis of $H_{A}$.

If we want to compute the time-behavior of $\langle O_{R}\rangle _{\rho (t)}$%
, we have to consider that $\Sigma ^{\left( 1\right) }$ is a sum of terms of
the form $(B_{\lambda ,\lambda ^{\prime }}\left\vert \alpha _{j}\right\vert
^{2}+B_{\lambda ,\lambda ^{\prime }}\left\vert \beta _{j}\right\vert ^{2})$,
that is, terms of the expectation value coming from the diagonal part of $%
\rho (t)$ in the basis of the Hamiltonian $H$. Therefore, if there is
decoherence, the sum $\Sigma ^{nd}(t)=\Sigma ^{\left( 2\right) }+\Sigma
^{\left( 3\right) }$, involving the terms of $\langle O_{R}\rangle _{\rho
(t)}$ coming from the non-diagonal part of $\rho (t)$, has to approach zero
for $t\rightarrow \infty $.

In order to show an example of the time-behavior of $\langle O_{R}\rangle
_{\rho (t)}$, numerical simulations for $\Sigma ^{nd}(t)$ have been
performed, with the following features:

\begin{enumerate}
\item[(i)] $s_{\lambda ^{\prime },\lambda }=1$ (see eq. (\ref{5.1})).

\item[(ii)] The initial condition for $S=A$ is selected as (see eq. (\ref%
{4.13})) : 
\begin{equation}
\left\vert \mathcal{A}\right\rangle =\frac{1}{\sqrt{2^{M}}}%
\sum_{i=1}^{2^{M}}\left\vert \mathcal{A}_{i}\right\rangle \Longrightarrow
\forall \lambda ,\ C_{\lambda }=C_{\lambda }^{\ast }=\frac{1}{\sqrt{2^{M}}}%
\Longrightarrow C_{\lambda }C_{\lambda ^{\prime }}^{\ast }=\frac{1}{2^{M}}
\end{equation}%
Then, from (i) and (ii), $B_{\lambda ,\lambda ^{\prime }}=2^{-M}$ (see eq. (%
\ref{4.25})).

\item[(iii)] $\left\vert \alpha _{i}\right\vert ^{2}$ is generated by a
random-number generator in the interval $[0,1]$, and $\left\vert \beta
_{i}\right\vert ^{2}$ is obtained as $\left\vert \beta _{i}\right\vert
^{2}=1-\left\vert \alpha _{i}\right\vert ^{2}$.

\item[(iv)] $g_{i}=400Hz$: as explained above, the coupling constant in
typical models of spin interaction.

\item[(v)] As in the original model, the time-interval $\left[ 0,t_{0}\right]
$ was partitioned into intervals $\Delta t=t_{0}/200$, and the function $%
\Sigma ^{nd}(t)$ was computed at times $t_{k}=k\Delta t$, with $%
k=0,1,...,200 $.

\item[(vi)] $N=10^{3}$, and $M=1$ and $M=10$.
\end{enumerate}

Figure 2 shows the time-evolution of $\Sigma ^{nd}(t)$.%

\begin{figure}[t]
 \centerline{\scalebox{0.7}{\includegraphics{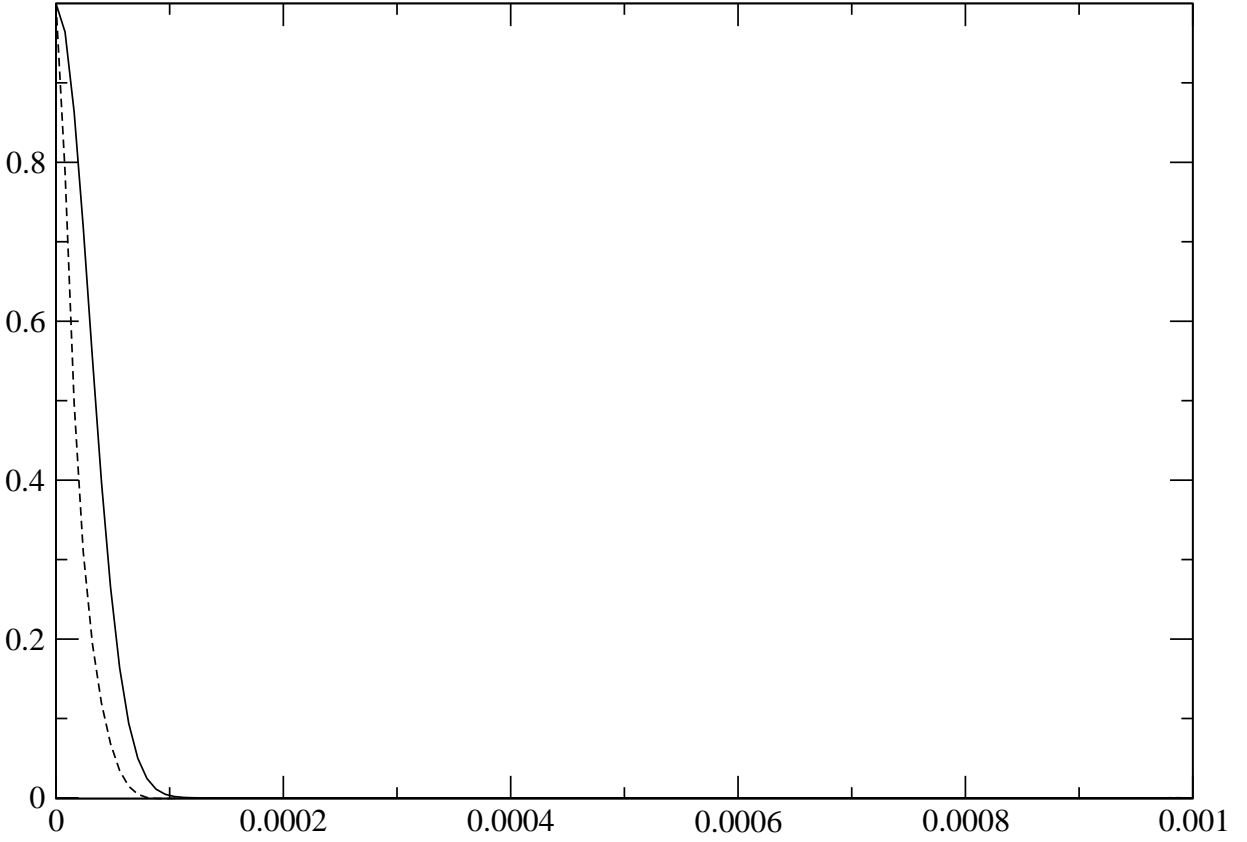}}}
\caption{Figure 2: Evolution of $%
\Sigma ^{nd}(t)$ for $N=10^{3}$, and $M=1$ (solid line) and $M=10$ (dash
line), with $t_{0}=10^{-3}s$.}
 \label{fig 2}\vspace*{0.cm}
\end{figure}

This result shows that, as expected, a small open system $S=A$ of $M$
particles decoheres in interaction with a large environment $E=B$ of $N\gg M$
particles.\bigskip

\textbf{Case (b): }$M\gg N$

In this case, where the open system $S=$ $A$ has much more particles than
the environment $E=B$, the argument of Case (a) cannot be applied: since now 
$\Sigma ^{\left( 2\right) }(t)$ and $\Sigma ^{\left( 3\right) }(t)$ are no
longer sums over a small number of terms, the fact that each term approaches
zero does not guarantee that the sums also approach zero. In particular, if $%
N=1$, then (see eq. (\ref{5.4}))%
\begin{equation}
T_{l,l^{\prime }}(t)=\left\vert \alpha _{1}\right\vert ^{2}e^{i\left(
g_{1,l}-g_{1,l^{\prime }}\right) \frac{t}{2}}+\left\vert \beta
_{1}\right\vert ^{2}e^{-i\left( g_{1,l}-g_{1,l^{\prime }}\right) \frac{t}{2}}
\label{5.16}
\end{equation}%
which clearly has no limit for $t\rightarrow \infty $. Nevertheless, it
might happen that, with $N$ high but $M$ much higher than $N$, each term of
the sums approaches zero. So, in order to know the time behavior of $\langle
O_{R}\rangle _{\rho (t)}$, numerical simulations for $\Sigma ^{nd}(t)$ have
been performed, with the same features as in the previous case, with the
exception of condition (vi), which was taken as:

\begin{enumerate}
\item[(vi)] $M=10^{3}$, and $N=10$ and $N=100$.
\end{enumerate}

Figure 3 shows the time-evolution of $\Sigma ^{nd}(t)$ in this case.%

\begin{figure}[t]
 \centerline{\scalebox{0.7}{\includegraphics{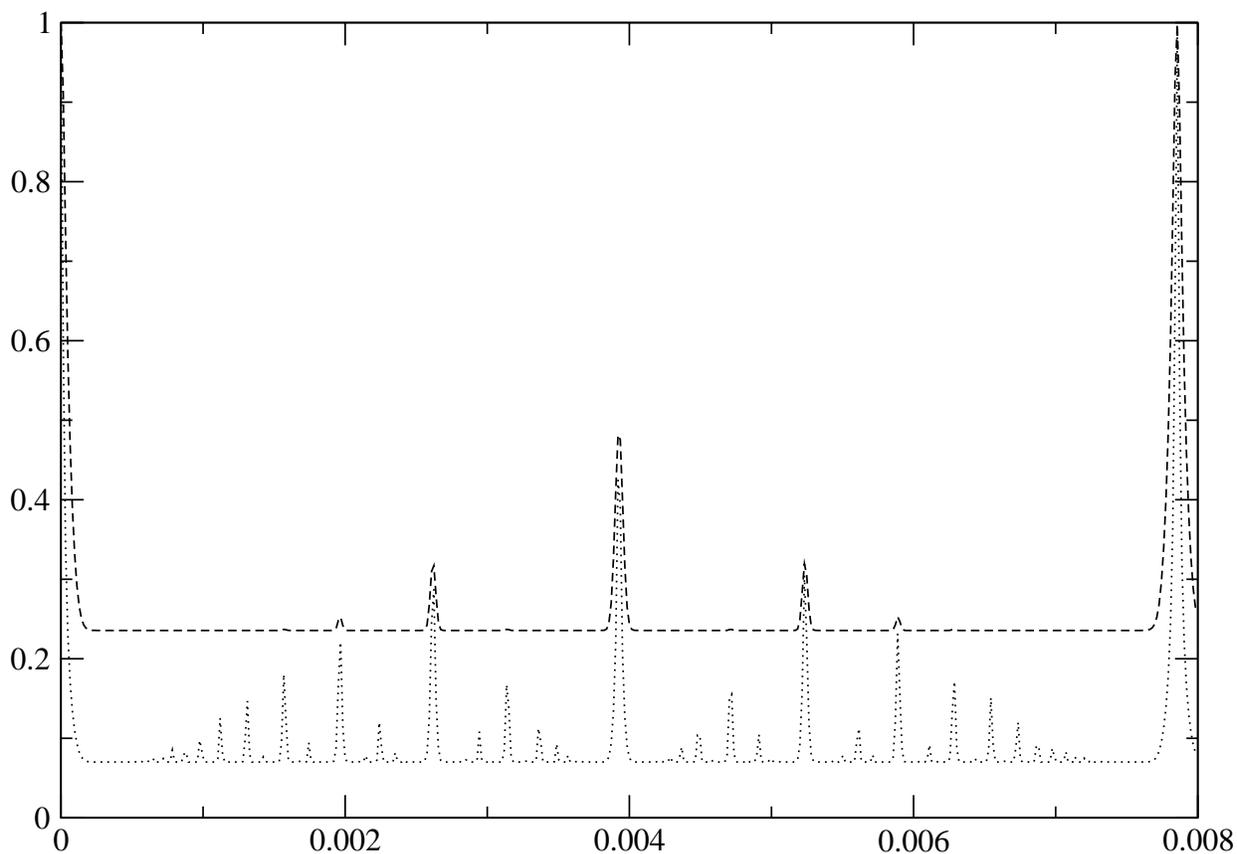}}}
\caption{Figure 3: Evolution of $%
\Sigma ^{nd}(t)$ for $M=10^{3}$, and $N=10$ (dash line) and $N=100$ (dot
line), with $t_{0}=10^{-3}s$.}
 \label{fig 3}\vspace*{0.cm}
\end{figure}

This result is also what may be expected: when \ the open system $S=A$ of $M$
particles is larger that the environment $E=B$ of $N\ll M$ particles, $S$
does not decohere.\bigskip

\textbf{Case (c): }$M\simeq N$

In this case, where the numbers of particles of the open system $S=$ $A$ and
of the environment $E=B$ do not differ in more than one order of magnitude,
the time behavior of $\langle O_{R}\rangle _{\rho (t)}$ cannot be inferred
from the equations. Numerical simulations have been performed, with the same
features as in Case (b), with the exception of condition (vi), which was
taken as:

\begin{enumerate}
\item[(vi)] $N=10^{3}$, and $M=10^{2}$ and $M=10^{3}$.
\end{enumerate}

Figure 4 shows the time-evolution of $\Sigma ^{nd}(t)$.%

\begin{figure}[t]
 \centerline{\scalebox{0.7}{\includegraphics{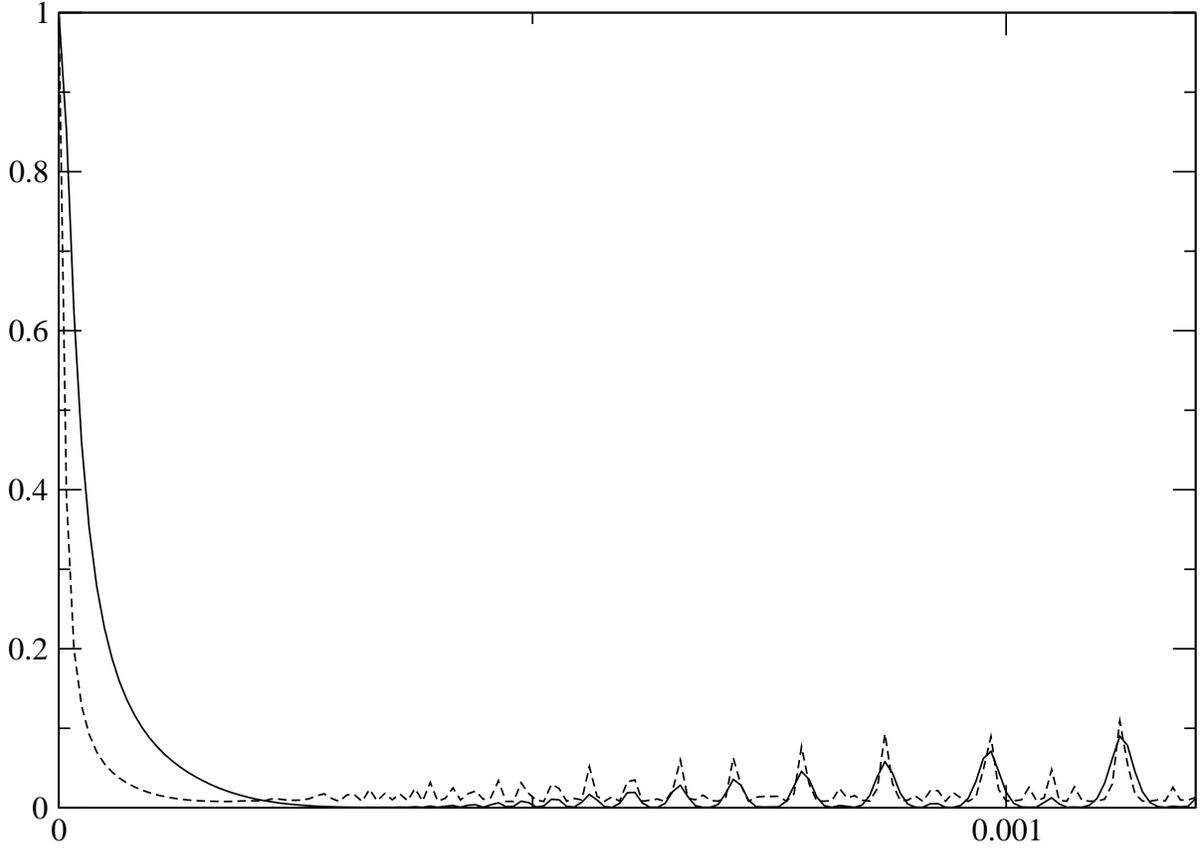}}}
\caption{Figure 4: Evolution of $%
\Sigma ^{nd}(t)$ for $N=10^{3}$, and $M=10^{2}$ (dash line) and $M=10^{3}$
(solid line), with $t_{0}=12.10^{-4}s$.}
 \label{fig 4}\vspace*{0.cm}
\end{figure}

Again, if the environment $E=B$ of $N$ particles is not large enough when
compared with the open system $S=A$ of $M$ particles, $S$ does not decohere.
Let us notice that, for $N=10^{3}$, the system $S=A$ with $M=10^{2}$ does
not decohere (Figure 4), whereas it does decohere with $M=10$ (Figure 2).
This shows that, in the case of this decomposition, $M\ll N$ means that $N$
is at least two orders of magnitude higher than $M$.\bigskip

\textbf{Summarizing results}

Up to now, in this Decomposition 1 all the arguments were directed to know
whether the system $A$ of $M$ particles decoheres or not in interaction with
the system $B$ of $N$ particles. But, given the symmetry of the whole
system, the same arguments can be used to decide whether the system $B$ of $%
N $ particles decoheres or not in interaction with the system $A$ of $M$
particles, with analogous results: $B$ decoheres only when $M\gg N$; if $%
M\ll N$ or $M\simeq N$, $B$ does not decohere. Therefore, all the results
obtained in this section can be summarized as follows:

\begin{enumerate}
\item[(i)] If $M\ll N$, $A$ decoheres and $B$ does not decohere.

\item[(ii)] If $M\gg N$, $A$ does not decohere and $B$ decoheres.

\item[(iii)] If $M\simeq N$, neither $A$ nor $B$ decohere.
\end{enumerate}

\section{Generalized spin-bath model: Decomposition 2}

\subsection{Selecting the relevant observables}

In this case we decide to observe only one particle of the open system $A$.
This amounts to splitting the closed system $U$ into two new subsystems: the
open system $S$ is, say, the particle $A_{M}$ with ket $\left\vert \Uparrow
,\Uparrow ,...,\Uparrow ,\Uparrow ,\Uparrow ,\Downarrow \right\rangle $, and
the environment is $E=\left( \cup _{i=1}^{M-1}A_{i}\right) \cup B=\left(
\cup _{i=1}^{M-1}A_{i}\right) \cup \left( \cup _{k=1}^{N}B_{k}\right) $. The
decomposition for this case is%
\begin{equation}
\mathcal{H}=\mathcal{H}_{S}\otimes \mathcal{H}_{E}=\left( \mathcal{H}%
_{A_{M}}\right) \otimes \left( \left( \bigotimes\limits_{i=1}^{M-1}\mathcal{H%
}_{A_{i}}\right) \otimes \left( \bigotimes\limits_{k=1}^{N}\mathcal{H}%
_{B_{k}}\right) \right)  \label{6.0}
\end{equation}%
Therefore, the relevant observables $O_{R}$ of the closed system $U$ are
those corresponding to the particle $A_{M}$:%
\begin{equation}
O_{R}=O_{S}\otimes \mathbb{I}_{E}=\left( \sum\limits_{\alpha ,\alpha
^{\prime }=\Uparrow ,\Downarrow }s_{\alpha ,\alpha ^{\prime }}\left\vert
\alpha \right\rangle \left\langle \alpha ^{\prime }\right\vert \right)
\otimes \left( \left( \bigotimes_{i=1}^{M-1}\mathbb{I}_{i}\right) \otimes
\left( \bigotimes_{k=1}^{N}\mathbb{I}_{k}\right) \right)  \label{6.1}
\end{equation}%
It is easy to see that the relevant observables selected in this
Decomposition 2 form a subspace of the space of the relevant observables
selected in Decomposition 1: eq. (\ref{6.1}) can be obtained from eq. (\ref%
{5.1}) by making $s_{\lambda ,\lambda ^{\prime }}=1$ for $\lambda =\lambda
^{\prime }$ and $s_{\lambda ,\lambda ^{\prime }}=0$ for $\lambda \neq
\lambda ^{\prime }$ in all the terms of the sum except for the terms
corresponding to the particle $A_{M}$.

In order to simplify expressions, in this case it is convenient to introduce
a new arrangement for the eigenvectors of the Hamiltonian $H_{A}$, by
calling them $\left\vert \mathcal{\tilde{A}}_{i}\right\rangle $: the set $%
\left\{ \left\vert \mathcal{\tilde{A}}_{i}\right\rangle \right\} $ is an
eigenbasis of $H_{A}$ with $2^{M}$ elements. The $\left\vert \mathcal{\tilde{%
A}}_{i}\right\rangle $ will be ordered by analogy with the binary numbers:%
\begin{eqnarray}
\left\vert \mathcal{\tilde{A}}_{1}\right\rangle &=&\left\vert \Uparrow
,\Uparrow ,...,\Uparrow ,\Uparrow ,\Uparrow ,\Uparrow \right\rangle \text{, }%
\left\vert \mathcal{\tilde{A}}_{2}\right\rangle =\left\vert \Uparrow
,\Uparrow ,...,\Uparrow ,\Uparrow ,\Uparrow ,\Downarrow \right\rangle \text{%
, }\left\vert \mathcal{\tilde{A}}_{3}\right\rangle =\left\vert \Uparrow
,\Uparrow ,...,\Uparrow ,\Uparrow ,\Downarrow ,\Uparrow \right\rangle \text{,%
}  \notag \\
\text{ }\left\vert \mathcal{\tilde{A}}_{4}\right\rangle &=&\left\vert
\Uparrow ,\Uparrow ,...,\Uparrow ,\Uparrow ,\Downarrow ,\Downarrow
\right\rangle \text{, }\left\vert \mathcal{\tilde{A}}_{5}\right\rangle
=\left\vert \Uparrow ,\Uparrow ,...,\Uparrow ,\Downarrow ,\Uparrow ,\Uparrow
\right\rangle \text{, }\left\vert \mathcal{\tilde{A}}_{6}\right\rangle
=\left\vert \Uparrow ,\Uparrow ,...,\Uparrow ,\Downarrow ,\Uparrow
,\Downarrow \right\rangle ,...  \notag \\
\left\vert \mathcal{\tilde{A}}_{2^{M}}\right\rangle &=&\left\vert \Downarrow
,\Downarrow ,...,\Downarrow ,\Downarrow ,\Downarrow ,\Downarrow \right\rangle
\label{6.2}
\end{eqnarray}%
According to this arrangement, the $\left\vert \mathcal{\tilde{A}}%
_{i}\right\rangle $ with even $i$ have the spin $M$ in the state $\left\vert
\Downarrow \right\rangle $, and the $\left\vert \mathcal{\tilde{A}}%
_{i}\right\rangle $ with odd $i$ have the spin $M$ in the state $\left\vert
\Uparrow \right\rangle $. So, the relevant observables of eq. (\ref{6.1})
can be rewritten in terms of the $\left\vert \mathcal{\tilde{A}}%
_{i}\right\rangle $ as%
\begin{equation}
O_{R}=\left( \sum\limits_{\lambda =1}^{2^{M}}\left( \tilde{s}_{\Uparrow
\Uparrow }\left\vert \mathcal{\tilde{A}}_{2\lambda }\right\rangle
\left\langle \mathcal{\tilde{A}}_{2\lambda }\right\vert +\tilde{s}_{\Uparrow
\Downarrow }\left\vert \mathcal{\tilde{A}}_{2\lambda }\right\rangle
\left\langle \mathcal{\tilde{A}}_{2\lambda -1}\right\vert +\tilde{s}%
_{\Downarrow \Uparrow }\left\vert \mathcal{\tilde{A}}_{2\lambda
-1}\right\rangle \left\langle \mathcal{\tilde{A}}_{2\lambda }\right\vert +%
\tilde{s}_{\Downarrow \Downarrow }\left\vert \mathcal{\tilde{A}}_{2\lambda
-1}\right\rangle \left\langle \mathcal{\tilde{A}}_{2\lambda -1}\right\vert
\right) \right) \otimes \left( \bigotimes_{k=1}^{N}\mathbb{I}_{k}\right)
\label{6.3}
\end{equation}

\subsection{Computing the behavior of the relevant expectation values}

Here the expectation values of the relevant observables are given by eq. (%
\ref{4.26}), with $T_{l,l^{\prime }}(t)$, $T_{l,l}(t)$ and $T_{l,M-l}(t)$
given by eqs. (\ref{4.24}), (\ref{4.28}) and (\ref{4.29}) respectively, but
now replacing $B_{\lambda ,\lambda ^{\prime }}$ with $\tilde{B}_{\lambda
,\lambda ^{\prime }}$, 
\begin{equation}
\langle O_{R}\rangle _{\rho (t)}=\sum\limits_{l=0}^{M}\sum\limits_{\QATOP{%
\lambda =f(l-1)+1}{\lambda ^{\prime }=f(l-1)+1}}^{f(l)}\tilde{B}_{\lambda
,\lambda ^{\prime }}+\sum\limits_{l=0}^{\tilde{M}}\sum\limits_{\QATOP{%
\lambda =f(l-1)+1}{\lambda ^{\prime }=f(M-l-1)+1}}^{\QATOP{f(l)}{f(M-l)}}%
\tilde{B}_{\lambda ,\lambda ^{\prime }}2\func{Re}\left( T_{l,M-l}(t)\right)
+\sum\limits_{\QATOP{\QATOP{l,l^{\prime }=0}{l\neq l^{\prime }}}{l^{\prime
}\neq M-l}}^{M}\sum\limits_{\QATOP{\lambda =f(l-1)+1}{\lambda ^{\prime
}=f(l^{\prime }-1)+1}}^{\QATOP{f(l)}{f(l^{\prime })}}\tilde{B}_{\lambda
,\lambda ^{\prime }}T_{l,l^{\prime }}(t)  \label{6.4}
\end{equation}%
where the $\tilde{B}_{\lambda ,\lambda ^{\prime }}$ can be written in the
basis $\left\{ \left\vert \mathcal{\tilde{A}}_{\lambda }\right\rangle
\right\} $ as%
\begin{equation}
\tilde{B}_{\lambda ,\lambda ^{\prime }}=\left\{ 
\begin{array}{ccccc}
C_{\lambda }C_{\lambda ^{\prime }}^{\ast }\tilde{s}_{\Uparrow \Uparrow } & 
\text{if} & \lambda & \text{is an even number and} & \lambda ^{\prime
}=\lambda \\ 
C_{\lambda }C_{\lambda ^{\prime }}^{\ast }\tilde{s}_{\Uparrow \Downarrow } & 
\text{if} & \lambda & \text{is an even number and} & \lambda ^{\prime
}=\lambda -1 \\ 
C_{\lambda }C_{\lambda ^{\prime }}^{\ast }\tilde{s}_{\Downarrow \Uparrow } & 
\text{if} & \lambda & \text{is an odd number and} & \lambda ^{\prime
}=\lambda +1 \\ 
C_{\lambda }C_{\lambda ^{\prime }}^{\ast }\tilde{s}_{\Downarrow \Downarrow }
& \text{if} & \lambda & \text{is an odd number and} & \lambda ^{\prime
}=\lambda \\ 
0 &  &  & \text{otherwise} & 
\end{array}%
\right\}  \label{6.5}
\end{equation}%
According to eq. (\ref{6.5}), $\tilde{B}_{\lambda ,\lambda ^{\prime }}\neq 0$
only when

\begin{equation}
\lambda ^{\prime }=\lambda \text{ \ \ \ \ or \ \ \ }\lambda ^{\prime
}=\lambda \pm 1\text{ }  \label{6.5b}
\end{equation}%
Since $\lambda =f(l-1)+1$ and $\lambda ^{\prime }=f(l^{\prime }-1)+1$,
relations (\ref{6.5b}) imply that 
\begin{equation}
l^{\prime }=l\text{ \ \ \ \ or \ \ \ }l^{\prime }=l\pm 1  \label{6.5c}
\end{equation}

The expectation value given by eq. (\ref{6.4}) has again three terms, $%
\langle O\rangle _{\rho (t)}=\Sigma ^{\left( 1\right) }+\Sigma ^{\left(
2\right) }+\Sigma ^{\left( 3\right) }$, which can be analyzed separately:

\begin{itemize}
\item From eqs. (\ref{6.5}) and (\ref{6.5b}), the first term reads%
\begin{equation}
\Sigma ^{\left( 1\right) }=\sum\limits_{l=0}^{M}\sum\limits_{\lambda
=f(l-1)+1}^{f(l)}B_{\lambda ,\lambda }=\sum\limits_{\lambda
=0}^{2^{M-1}}\left( \left\vert C_{2\lambda }\right\vert ^{2}\tilde{s}%
_{\Uparrow \Uparrow }+\left\vert C_{2\lambda +1}\right\vert ^{2}\tilde{s}%
_{\Downarrow \Downarrow }\right) \neq \Sigma ^{\left( 1\right) }(t)
\label{6.5d}
\end{equation}%
Analogously to eq. (\ref{5.7}) of Decomposition 1, this first term does not
evolve with time.

\item The time-dependence of the second term is given by $T_{l,M-l}(t)$. But
with the restrictions of eqs. (\ref{6.5b}) and (\ref{6.5c}), $\Sigma
^{\left( 2\right) }$ has only two terms:%
\begin{eqnarray}
\Sigma ^{\left( 2\right) }(t) &=&\sum\limits_{l=0}^{\tilde{M}}\sum\limits_{%
\QATOP{\lambda =f(l-1)+1}{\lambda ^{\prime }=f(M-l-1)+1}}^{\QATOP{f(l)}{%
f(M-l)}}B_{\lambda ,\lambda ^{\prime }}2\func{Re}\left( T_{l,M-l}(t)\right) =
\\
&=&C_{f(\frac{M-1}{2}-1)+1}C_{f(\frac{M-1}{2}-1)+2}^{\ast }\left( \tilde{s}%
_{\Downarrow \Uparrow }+\tilde{s}_{\Uparrow \Downarrow }\right) 2\func{Re}%
\left( T_{\frac{M-1}{2},\frac{M+1}{2}}(t)\right)  \label{6.5.e}
\end{eqnarray}%
Then, in order to obtain the limit of this term, we have to compute the
limit of $T_{\frac{M-1}{2},\frac{M+1}{2}}(t)$, which is precisely the $%
T_{l,l^{\prime }}(t)$ of Decomposition 1 in the particular case that $l=%
\frac{M-1}{2}$ and $l^{\prime }=\frac{M+1}{2}$ (see eq. (\ref{5.4})). But,
as we have seen in Case (a) of Decomposition 1, $T_{l,l^{\prime }}(t)$ has
the same functional form as $\Gamma _{1}(t)$ of the original model (see eq. (%
\ref{3.10})), which approaches zero for $t\rightarrow \infty $ when $N\gg 1$%
. Therefore, for $N\gg 1$, $T_{\frac{M-1}{2},\frac{M+1}{2}}(t)$ also
approaches zero for $t\rightarrow \infty $, and the same holds for $\Sigma
^{\left( 2\right) }(t)$ since it is a sum of two terms containing $T_{\frac{%
M-1}{2},\frac{M+1}{2}}(t)$.

\item The time-dependence of the third term is given by $T_{l,l^{\prime
}}(t) $. But with the restrictions of eqs. (\ref{6.5b}) and (\ref{6.5c}), $%
\Sigma ^{\left( 3\right) }$ results:%
\begin{equation}
\Sigma ^{\left( 3\right) }(t)=\sum\limits_{\QATOP{l=0}{l\neq \frac{M-1}{2}}%
}^{M}\sum\limits_{\lambda =f(l-1)+1}^{f(l)}\left( B_{\lambda ,\lambda
+1}T_{l,l+1}(t)+B_{\lambda ,\lambda -1}T_{l,l-1}(t)\right)  \label{6.5h}
\end{equation}%
Since here $l^{\prime }=l\pm 1$ (see eq. (\ref{6.5c})), in this case $%
T_{l,l\pm 1}(t)$ is:%
\begin{equation}
T_{l,l\pm 1}(t)=\prod_{j=1}^{N}\left( \left\vert \alpha _{j}\right\vert
^{2}e^{\mp ig_{j}t}+\left\vert \beta _{j}\right\vert ^{2}e^{\pm
ig_{j}t}\right)  \label{6.5i}
\end{equation}%
If we compare this equation with eq. (\ref{3.13}) for $r(t)$ in the original
spin-bath model, we can see that 
\begin{equation}
T_{l,l+1}(t)=r(t)\text{ \ \ \ \ and \ \ \ }T_{l,l-1}(t)=r^{\ast }(t)
\label{6.5j}
\end{equation}%
Then, 
\begin{equation}
\Sigma ^{\left( 3\right) }(t)=\left( S_{+}r(t)+S_{-}r^{\ast }(t)\right)
\label{6.5k}
\end{equation}%
where $S_{+}$ and $S_{-}$ are constants given by 
\begin{equation}
S_{\pm }=\sum\limits_{\QATOP{l=0}{l\neq \frac{M-1}{2}}}^{M}\sum\limits_{%
\lambda =f(l-1)+1}^{f(l)}B_{\lambda ,\lambda \pm 1}  \label{6.5m}
\end{equation}%
On the basis of the simulations of the original model we have seen that,
when $N\gg 1$, $r(t)$ approaches zero for $t\rightarrow \infty $. Therefore,
in this case we can conclude that, when $N\gg 1$, $\Sigma ^{\left( 3\right)
}(t)$ approaches zero for $t\rightarrow \infty $.
\end{itemize}

Summing up, $\langle O_{R}\rangle _{\rho (t)}$ is the sum of three terms:
one is time-independent and the other two tend to zero for $t\rightarrow
\infty $. In particular, from eq. (\ref{6.5d}) we know that, for $N\gg 1$,

\begin{equation}
\lim_{t\rightarrow \infty }\langle O_{R}\rangle _{\rho
(t)}=\sum\limits_{l=0}^{M}\sum\limits_{\lambda ,\lambda ^{\prime
}=f(l-1)+1}^{f(l)}\tilde{B}_{\lambda ,\lambda ^{\prime
}}=\sum\limits_{l=0}^{M}\sum\limits_{\lambda =0}^{2^{M-1}}\left( \left\vert
C_{2\lambda }\right\vert ^{2}\tilde{s}_{\Uparrow \Uparrow }+\left\vert
C_{2\lambda +1}\right\vert ^{2}\tilde{s}_{\Downarrow \Downarrow }\right)
=\langle O_{R}\rangle _{\rho _{\ast }}  \label{6.6}
\end{equation}%
where $\rho _{\ast }$ is the final diagonal state of $U$. Again, this result
can also be expressed in terms of the reduced density operator $\rho
_{S}=\rho _{A_{M}}$ of the open system $S=A_{M}$ as (see eq. (\ref{5.13}))%
\begin{equation}
\lim_{t\rightarrow \infty }\langle O_{R}\rangle _{\rho (t)}=\langle
O_{R}\rangle _{\rho _{\ast }}=\lim_{t\rightarrow \infty }\langle
O_{A_{M}}\rangle _{\rho _{A_{M}}(t)}=\langle O_{A_{M}}\rangle _{\rho
_{A_{M}\ast }}  \label{6.7}
\end{equation}%
where the final reduced density operator $\rho _{A_{M}\ast }$ in the basis $%
\{\left\vert \Uparrow \right\rangle ,\left\vert \Downarrow \right\rangle \}$
reads%
\begin{equation}
\rho _{A_{M}\ast }=%
\begin{pmatrix}
\left\vert \alpha _{M}\right\vert ^{2} & 0 \\ 
0 & \left\vert \beta _{M}\right\vert ^{2}%
\end{pmatrix}
\label{6.8}
\end{equation}%
This shows that the open system $S=A_{M}$, composed of a single particle,
decoheres in interaction with its environment $E$ of $N+M-1$ particles when $%
N\gg 1$, independently of the value of $M$.\bigskip

In order to illustrate this conclusion, we have computed $\Sigma
^{nd}(t)=\Sigma ^{\left( 2\right) }(t)+\Sigma ^{\left( 3\right) }(t)$ by
means of numerical simulations with the same features as in Decomposition 1,
with the exception of condition (vi), which was taken as:\medskip

Figure 5: $\ $(vi) $\ M=10^{3}$ and $N=1$.\medskip

Figure 6: $\ $(vi) $\ M=10^{3}$ and $N=10^{2}$.\medskip

Figure 7: $\ $(vi) $\ M=10^{3}$ and $N=10^{3}$.\medskip

\begin{figure}[t]
 \centerline{\scalebox{0.7}{\includegraphics{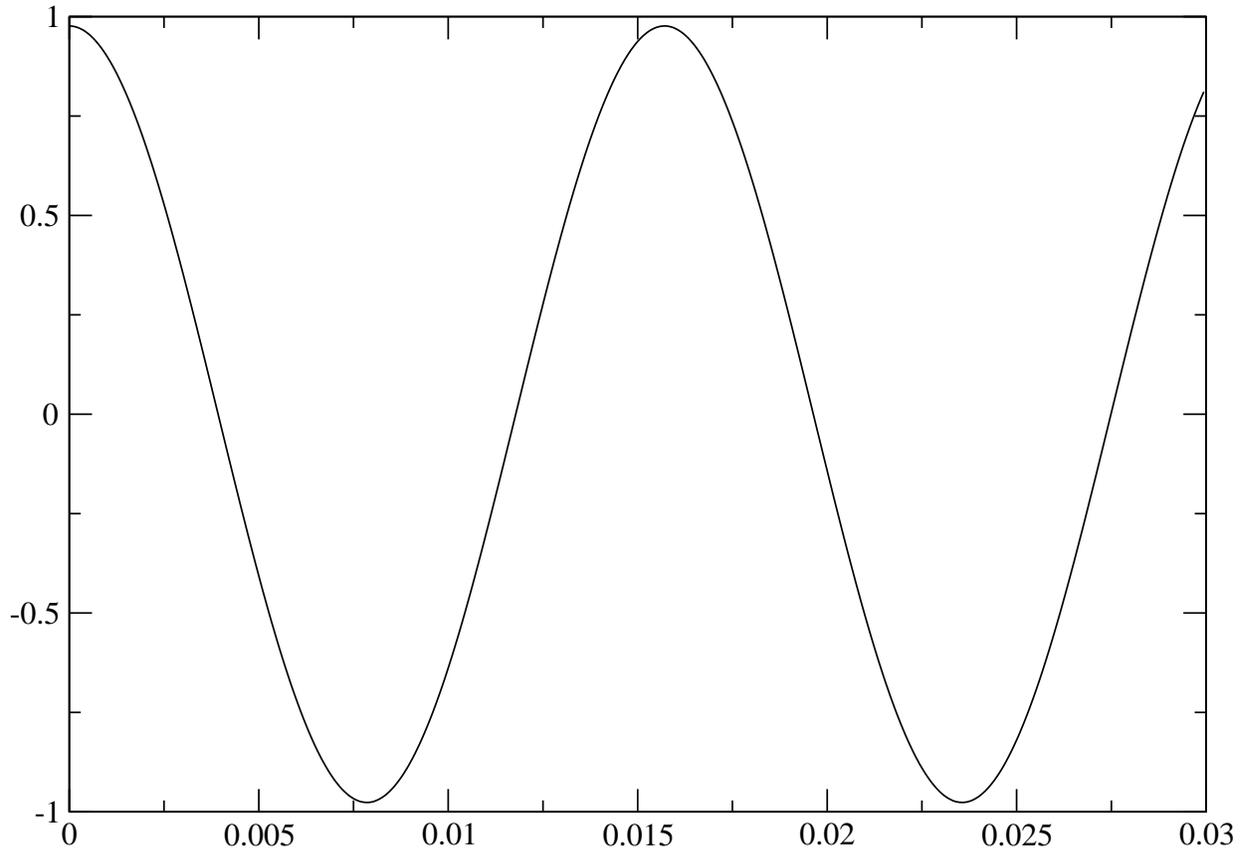}}}
\caption{Figure 5:
Evolution of $\Sigma ^{nd}(t)$ for $M=10^{3}$ and $N=1$, with $%
t_{0}=3.10^{-2}s$.}
 \label{fig 5}\vspace*{0.cm}
\end{figure}

\begin{figure}[t]
 \centerline{\scalebox{0.7}{\includegraphics{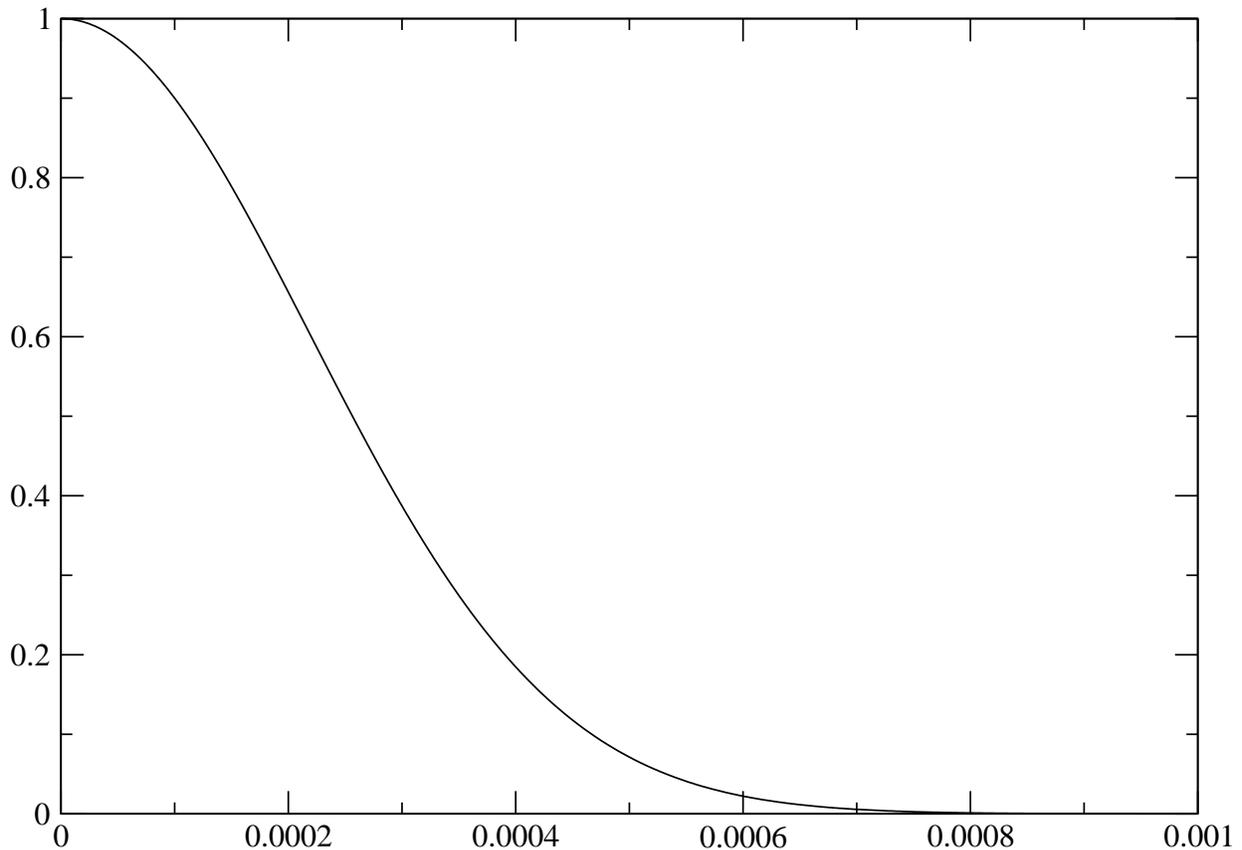}}}
\caption{Figure 6: Evolution of $%
\Sigma ^{nd}(t)$ for $M=10^{3}$ and $N=10^{2}$, with $t_{0}=1.10^{-3}s$.}
 \label{fig 6}\vspace*{0.cm}
\end{figure}

\begin{figure}[t]
 \centerline{\scalebox{0.7}{\includegraphics{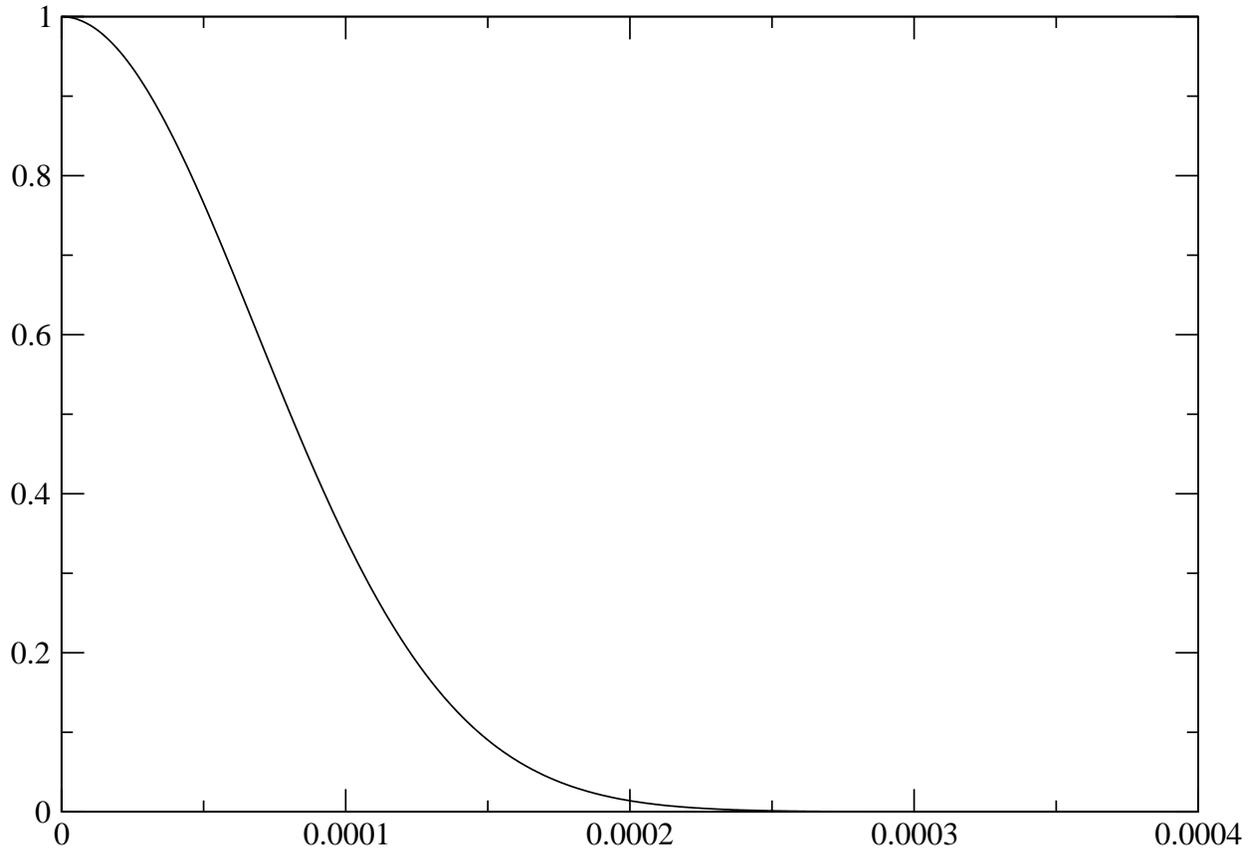}}}
\caption{Figure 7: Evolution of $%
\Sigma ^{nd}(t)$ for $M=10^{3}$ and $N=10^{3}$, with $t_{0}=4.10^{-4}s$.}
 \label{fig 7}\vspace*{0.cm}
\end{figure}

\textbf{Summarizing results}

As we have seen, in this decomposition of the whole closed system, the open
system $S=A_{M}$ decoheres when $N\gg 1$, independently of the value of $M$.
But the particle $A_{M}$ was selected as $S$ only for computation
simplicity: the same argument can be developed for any particle $A_{i}$ of $%
A $. Then, when $N\gg 1$ and independently of the value of $M$, any particle 
$A_{i}$ decoheres in interaction with its environment $E$ of $N+M-1$
particles.

On the other hand, as in Decomposition 1, here the symmetry of the whole
system $U$ allows us to draw analogous conclusions when the system $S$ is
one of the particles of $B$, say, $B_{N}$: $S=B_{N}$ decoheres when $M\gg 1$%
, independently of the value of $N$. And, on the basis of the same
considerations as above, when $M\gg 1$ and independently of the value of $N$%
, any particle $B_{i}$ decoheres in interaction with its environment $E$ of $%
N+M-1$ particles.

\section{Concluding remarks}

In this paper we have studied a generalization of the spin-bath model, where
a closed system $U$ is composed by two subsystems, $U=A\cup B$, with $A$ of $%
M$ particles $A_{i}$ and $B$ of $N$ particles $B_{i}$. We showed how the
model behaves under different definitions of the system of interest and
under different relations between the numbers $M$ and $N$. \ The results so
obtained allow us to state the following concluding remarks:

\begin{enumerate}
\item[a)] We have seen that, when $M\gg N$ or $M\simeq N$, the subsystem $A$
does not decohere (Decomposition 1 of Section IV), but the particles $A_{i}$%
, considered independently, decohere when $N\gg 1$ (Decomposition 2 of
Section V). This means that there are physically meaningful situations,
given by $M\gg N\gg 1$ or $M\simeq N\gg 1$, where all the $A_{i}$ decohere
although $A$ does not decohere. In other words, in spite of the fact that
certain particles decohere and may behave classically, the subsystem
composed by all of them retains its quantum nature. We have also seen that,
by symmetry, all the particles $B_{i}$, considered independently, also
decohere when $M\gg 1$. Then, when $M\gg N\gg 1$ or $M\simeq N\gg 1$, the
requirement $M\gg 1$ holds and we can conclude that not only all the $A_{i}$%
, but also all the $B_{i}$ decohere, although $B$ neither decoheres. So, all
the particles of the closed system $U=$ $\left( \cup _{i}A_{i}\right) \cup
\left( \cup _{j}B_{j}\right) $ may become classical when considered
independently, although the whole system $U$ certainly does not decohere
and, therefore, retains its quantum character. These results, considered
together, are a clear manifestation of the fact, already pointed out by
Schlosshauer (\cite{Schlosshauer}), that energy dissipation and decoherence
are different phenomena: since all the particles of the system $U$ decohere
when independently considered, decoherence cannot result from the
dissipation of energy from the decohered systems to their environments.

\item[b)] The generalized model shows that the split of the entire closed
system into an open system and its environment amounts to the selection of
the observables relevant in each situation. Since there is no privileged or
essential decomposition, we can select the observables of the subsystem $A$
in the situation in which $A$ \ does not decohere. In this way, it would be
possible to use appropriately selected subsystems, unaffected by
decoherence, for storing quantum information.

\item[c)] The natural further step of generalization will consist in
following the ideas of paper \cite{Tessieri}, and introducing coupling
internal to the subsystems $A$ or $B$. For instance, given that the
decoherence of $A$ is increasingly suppressed as the number $M$ of its
particles increases, it could be expected that such decoherence suppression
will also be more efficient as the interactions between the spins of the
bath also increase.
\end{enumerate}

\end{document}